\documentclass[preprint,showpacs,showkeys,preprintnumbers,amsmath,amssymb,aps,groupedaddress]{revtex4}

\usepackage{graphicx}       
\usepackage{psfrag}
\usepackage{tikz}
\usepackage{mathptmx}
\usepackage{amsmath,amscd}
\usetikzlibrary{arrows,shapes,backgrounds}

\newcommand{\const}{\mathop{\mathrm{const}}}

\begin{document}

\title{Cracking the Taub-{NUT}}

\author{Pierre-Philippe Dechant}
\email{p.dechant@mrao.cam.ac.uk}
\affiliation{Astrophysics Group, Cavendish Laboratory, J J Thomson Avenue,  Cambridge, CB3 0HE, UK, and Kavli Institute for Cosmology, Cambridge}

\author{Anthony N. Lasenby}
\email{a.n.lasenby@mrao.cam.ac.uk}
\affiliation{Astrophysics Group, Cavendish Laboratory, J J Thomson Avenue, Cambridge, CB3 0HE, UK, and Kavli Institute for Cosmology, Cambridge}

\author{Michael P. Hobson} \email{mph@mrao.cam.ac.uk}
\affiliation{Astrophysics Group, Cavendish Laboratory, J J Thomson
  Avenue, Cambridge, CB3 0HE, UK}

\date{\today}

\begin{abstract}
We present further analysis of an anisotropic, non-singular early
universe model that leads to the viable cosmology presented in
\cite{DechantHobsonLasenby2009BianchiIXScalarField}.  Although this
model (the DLH model) contains scalar field matter, it is reminiscent
of the Taub-NUT vacuum solution in that it has biaxial Bianchi IX
geometry and its evolution exhibits a dimensionality reduction at a
quasi-regular singularity that one can identify with the big-bang.  We
show that the DLH and Taub-NUT metrics are related by a
coordinate transformation, in which the DLH time coordinate plays the
role of conformal time for Taub-NUT. Since both models continue
through the big-bang, the coordinate transformation can become
multivalued.  In particular, in mapping from DLH to Taub-NUT, the
Taub-NUT time can take only positive values.  We present explicit maps
between the DLH and Taub-NUT models, with and without a scalar
field. In the vacuum DLH model, we find a periodic solution
expressible in terms of elliptic integrals; this periodicity is broken
in a natural manner as a scalar field is gradually introduced to
recover the original DLH model. Mapping the vacuum solution over to
Taub-NUT coordinates, recovers the standard (non-periodic) Taub-NUT
solution in the Taub region, where Taub-NUT time takes positive
values, but does not exhibit the two NUT regions known in the standard
Taub-NUT solution. Conversely, mapping the complete Taub-NUT solution
to the DLH case reveals that the NUT regions correspond to imaginary
time and space in DLH coordinates.  We show that many of the
well-known `pathologies' of the Taub-NUT solution arise because the
traditional coordinates are connected by a multivalued transformation
to the physically more meaningful DLH coordinates. In particular, the
`open-to-closed-to-open' transition and the Taub and NUT regions of
the (Lorentzian) Taub-NUT model are replaced by a closed pancaking
universe with spacelike homogeneous  sections at all times.
\end{abstract}

\pacs{98.80.Bp 
, 98.80.Cq 
, 98.80.Jk 
, 04.20.Dw 
, 04.20.Jb 
, 04.20.dc
}

\keywords{scalar fields, Bianchi models, big bang singularity, 
cosmology, exact solutions, Taub-NUT, pre-Big-Bang scenarios}

\maketitle
\tableofcontents

\section{Introduction\label{introduction}}

In a previous work \cite{DechantHobsonLasenby2009BianchiIXScalarField}
we argued that it is natural to consider a generalisation of the
standard cosmological scenario to one in which a scalar field
dominates the dynamics of a homogeneous but, in general, anisotropic
(Bianchi) universe. We presented a new solution (the DLH model) to the
cosmological field equations based on a closed biaxial Bianchi IX
universe containing scalar field matter. This led to a nonsingular
`pancaking' model in which the spatial hypersurface volume goes to
zero instantaneously at the `big-bang', but all physical quantities,
such as curvature invariants and the matter energy density remain
finite, and continue smoothly through the big-bang. Moreover, we
showed that the model leads to a viable cosmology at late times,
exhibiting desirable features such as isotropisation and inflation, as
well as producing perturbation spectra consistent with observations.

We also noted in \cite{DechantHobsonLasenby2009BianchiIXScalarField}
that, despite containing scalar field matter, our model was
reminiscent of the Taub-NUT vacuum solution, since both have biaxial
Bianchi IX geometry and an evolution that exhibits a dimensionality
reduction at a quasi-regular singularity.  In this paper, we show that
the metrics for the DLH and Taub-NUT models are, in fact, related by a
coordinate transformation.  It is thus of interest to investigate the
explicit mapping between models based on each metric, with and without
scalar field matter. Moreover, we investigate the well-known
`pathologies' of the Taub-NUT solution, in the context of the mapping
to the DLH model. We contend that the natural coordinatisation of the
DLH model is more physical than the traditional coordinates used to
describe Taub-NUT. We thus consider the possibility that pathologies
such as the `open-to-closed-to-open' transition and the Taub and NUT
regions of the (Lorentzian) Taub-NUT model arise because the
traditional coordinates are connected by a singular transformation to
the physically more meaningful DLH coordinates.

This paper is organised as follows. We begin with a description of
Bianchi universes in Section \ref{maths}.  We then briefly review the
DLH model in Section \ref{DLH} and the Taub-NUT model in Section
\ref{TaubNUT}. We show that the two metrics are related by a
coordinate transformation in Section \ref{transformation}, where the
mapping naturally leads to a new reparameterised form of the Taub-NUT
metric. We investigate this reparameterised Taub-NUT model in Section
\ref{repTNUT} and find that it admits a simple scaling family of
solutions, in contrast to the conventional Taub-NUT setup. In Section~
\ref{scenarios}, we consider the mapping between the DLH and
reparameterised Taub-NUT models, with and without a scalar field, 
and interpret the solutions physically, before concluding in
Section \ref{conclusion}.

\section{Bianchi Models\label{maths}}

Bianchi universes are spatially homogeneous and therefore have a
3-dimensional group of isometries $G_{3}$ acting simply transitively
on spacelike hypersurfaces. The standard classification hence follows
Bianchi's classification of 3-parameter Lie groups
\cite{bianchi1897}.

We adopt the metric convention $\left(+ - - \,-\right)$. Roman letters
$a,b,c...$ from the beginning of the alphabet denote Lie algebra
indices.  Greek letters $\mu,\nu,\sigma...$ label spacetime indices,
whereas Roman letters $i,j,k...$ from the middle of the alphabet label
purely spatial ones.

The isometry group of a manifold is a Lie group $G$ and can be thought
of as infinitesimally generated by the Killing vectors $\xi$, which
obey
$\left[\xi_{\mu},\xi_{\nu}\right]=C_{\,\,\mu\nu}^{\sigma}\xi_{\sigma}$
where the $C_{\,\,\mu\nu}^{\sigma}$ are the structure constants of
$G$. These can be used to construct an invariant basis, which is often
useful to make the symmetry manifest.  This is a set of (basis) vector
fields $X_{\mu}$, each of which is invariant under $G$, i.e. has
vanishing Lie derivative with respect to all the Killing vectors such
that
\begin{equation}
\left[\xi_{\mu},X_{\nu}\right]=0.\label{KVinv}
\end{equation}
Such a basis can be constructed simply by imposing this relation at a
point for some chosen set of independent vector fields and using the
Killing vectors to drag them out across the manifold. The
integrability condition for this set of first-order differential
equations in fact amounts to demanding that the $C_{\,\,\mu\nu}^{\sigma}$ be
the structure constants of some group. The invariant vector fields satisfy
\begin{equation}
\left[X_{\mu},X_{\nu}\right]=-C_{\,\,\mu\nu}^{\sigma}X_{\sigma}.
\label{invbas}
\end{equation}
Denoting the duals of the $X_{\mu}$ (the so-called invariant 1-forms, or Maurer-Cartan forms) by $\omega^{\mu}$, the
corresponding curl relations for the dual basis are
\begin{equation}
d\omega^{\mu}=\frac{1}{2}C_{\,\,\sigma\tau}^{\mu}\omega^{\sigma}\wedge\omega^{\tau}.
\label{curl}
\end{equation}
Because the $X_{\mu}$ are invariant vectors, the metric can now be
expressed as
\begin{equation}
ds^{2}=g_{\mu\nu}\omega^{\mu}\omega^{\nu},\label{constmet}
\end{equation}
for some $g_{\mu\nu}$.

Bianchi models can be constructed in various different ways. For
simplicity, we use the fact that the timelike  vector generating
the foliation of spacetime into homogeneous spacelike hypersurfaces commutes
with the three Killing vectors within the hypersurfaces (generating
the homogeneity), and hence we choose a representation that is
diagonal:
\begin{equation}
   ds^{2}=dt^{2}-\gamma_{ij}(t)\omega^{i}\omega^{j}=dt^{2}-\gamma_{kl}(t)(e_{i}^{k}(x)dx^{i})(e_{j}^{l}(x)dx^{j}),
\label{metric}
\end{equation}
in terms of an explicit $(x,y,z)$ coordinate system.

As outlined in \cite{DechantHobsonLasenby2009BianchiIXScalarField},
the Bianchi classification of $G_{3}$ group types hinges on the
decomposition into irreducible parts of the spatial part of the
structure constants $C_{\,\, ij}^{k}$.  Imposing the Jacobi identity
then essentially leaves nine distinct choices of parameterisations
(zeroes and signs) of the structure constants corresponding to nine
different groups, called Bianchi I through Bianchi IX.  All Bianchi
models have a timelike  vector, which generates the preferred
foliation into spacelike hypersurfaces, and three spacelike Killing vectors,
generating the homogeneity on those hypersurfaces.  Both models that
we will consider, DLH and Taub-NUT, are Bianchi IX such that the
symmetry algebra is $\mathfrak{so}(3)\sim \mathfrak{su}(2)$.  

General Bianchi IX models are thought, generically, to exhibit complicated
dynamics and chaos.  The dynamics near the initial singularity of
vacuum and orthogonal perfect fluid models is believed to be governed
by Bianchi I and II vacuum states via the Kasner map (`Mixmaster
attractor'). This description in terms of successive Kasner periods
can be reformulated in terms of reflections and is called
`cosmological billiard motion', also known as `BKL analysis' after
Belinskii, Khalatnikov and Lifshitz
\cite{BelinskyKhalatnikovLifshitz1972,BelinskyKhalatnikovLifshitzKL1969,
  Misner1969Mixmaster,Novikov1973ProcessesNearSingularities,
  Ringstrom2001BianchiIXattractor,UgglaHeinzle2009mixmaster,UgglaHeinzle2009newproof}.
This turns out to be just an example of a more general phenomenon when
one considers {(super-)gravity} close to a spacelike singularity (the
`BKL-limit'). In this limit the gravitational theory can be recast in
terms of billiard motion in a portion of hyperbolic space, as above,
such that the dynamics is determined by successive reflections. These
reflections, however, are precisely the elements of a Lorentzian
Coxeter group, which are themselves the Weyl groups of
(infinite-dimensional) Kac-Moody algebras. This then leads to the
conjecture that these Kac-Moody algebras are in fact symmetries of the
underlying gravitational theory
\cite{Henneaux2008KacMoodyBKL,HennauxPersson2008SpacelikeSingularitiesAndHiddenSymmetriesofGravity,
  HenneauxPerssonWesley2008CoxeterGroupStructure}. There are also some
concerns about the discrete nature of the Kasner map, and a continuous
generalisation -- see, for instance
\cite{CornishLevin1997unambiguously, CornishLevin1997MixmasterChaotic,
  CornishLevin1997Fareytale}. For recent work on locally rotationally 
symmetric (LRS) Bianchi cosmologies with
anisotropic matter see, for example, \cite{Heinzle2009BianchiLRS}.
In \cite{DechantHobsonLasenby2009BianchiIXScalarField}, we remarked on 
how Bianchi models can be considered as a deformation of Friedmann-Robertson-Walker (FRW)
models, and observed how these perturbations freeze out during inflation. In fact, 
this can be understood in terms of the characterisation of Bianchi IX
models  as an FRW model deformed by long range gravitational waves 
\cite{Grishchuk1976Bianchi}.

For the DLH and Taub-NUT models that we will consider, however, there
exists an additional biaxial symmetry: two of the left-invariant
$SU(2)$ one-forms appear with the same coefficient in the metric. Thus
there is an additional right action by a $U(1)$ factor inside the
$SU(2)$ which acts by isometries, so we are considering a class of
metrics admitting an $SU(2)\times U(1)$ symmetry group. As shown in
\cite{DechantHobsonLasenby2009BianchiIXScalarField}, and demonstrated
further below, this additional symmetry allows for much simpler
dynamical evolution than in the full triaxial Bianchi IX case.

\section{The DLH model\label{DLH}}

For the DLH model, the metric of the form (\ref{metric}) is
\begin{equation}
ds^2=dt^2-{\textstyle\frac{1}{4}} R_1^2 (\omega^1)^2-{\textstyle\frac{1}{4}}R_2^2\left[(\omega^2)^2+(\omega^3)^2\right]
\label{lineelmtDLH2},
\end{equation}
which trivially reduces to FRW form in the special case where the two
scale factors are equal, $R_1(t)=R_2(t)$.  Following
\cite{DechantHobsonLasenby2009BianchiIXScalarField}, and with the usual definitions for the Hubble parameters
$H_{i}(t)\equiv{\dot{R_{i}}}/{R_{i}}$
for the different directions, the Einstein
field equations are easily computed and give two dynamical equations
for the two independent radii $R_1$ and $R_2$ (from now on we will
drop the explicit $t$ dependence of variables):
\begin{equation}
2\dot{H}_{2}+3H_{2}^{2}+\kappa p-\Lambda=\frac{1}{R_{2}^{2}}\left(3\frac{R_{1}^{2}}{R_{2}^{2}}-4\right)\label{special1}\end{equation}
and
\begin{equation}
2H_{1}^{2}+2\dot{H}_{1}-H_{2}^{2}+2H_{1}H_{2}+\kappa p-\Lambda=-\frac{1}{R_{2}^{2}}\left(5\frac{R_{1}^{2}}{R_{2}^{2}}-4\right),\label{special2}\end{equation}
as well as the Friedmann equation (or Hamiltonian constraint)
\begin{equation}
H_{2}^{2}+2H_{1}H_{2}-\kappa\rho-\Lambda=\frac{1}{R_{2}^{2}}
\left(\frac{R_{1}^{2}}{R_{2}^{2}}-4\right),\label{specialFr}\end{equation}
and equation of motion for a simple massive scalar field (for which the potential is given by $V(\phi)=\frac{1}{2}m^2\phi^2$)
\begin{equation}
  m^{2}\phi+\left(H_{1}+2H_{2}\right)\dot{\phi}+\ddot{\phi}=0.
\label{specialEoM}
\end{equation}

It turns out that these equations have relatively straightforward
series solutions in $t-t_0$. We choose $t_0=0$ for simplicity, and take
it to denote a big bang-like event.
In \cite{DechantHobsonLasenby2009BianchiIXScalarField} we have
presented two solutions with definite parity -- one even (bouncing)
and one odd (pancaking) in the non-degenerate scale factor $R_1$. Here
we are interested in the pancaking solution
\begin{align}
R_{1}(t)&=t\left(a_{0}+a_{2}t^{2}+a_{4}t^{4}+\dots\right)\notag\\
R_{2}(t)=R_{3}(t)&=b_{0}+b_{2}t^{2}+b_{4}t^{4}+\dots \,\,\,\,, \label{seriesansatz}\\
\phi(t)&=f_{0}+f_{2}t^{2}+f_{4}t^{4}+\dots\notag
\end{align}
where the dynamical equations (\ref{special1}), (\ref{special2}) and
(\ref{specialEoM}) allow one to fix the higher-order coefficients in
the series order-by-order in terms of the initial values
$a_0=\dot{R}_1(0)$, $b_0=R_2(0)$ and $f_0=\phi(0)$. The fact that this
also satisfies the Friedmann energy constraint (\ref{specialFr}) then
proves that this odd-parity series solution is a valid expansion
around the big-bang at $t=0$, which one can use as a starting point
for numerical integration. 

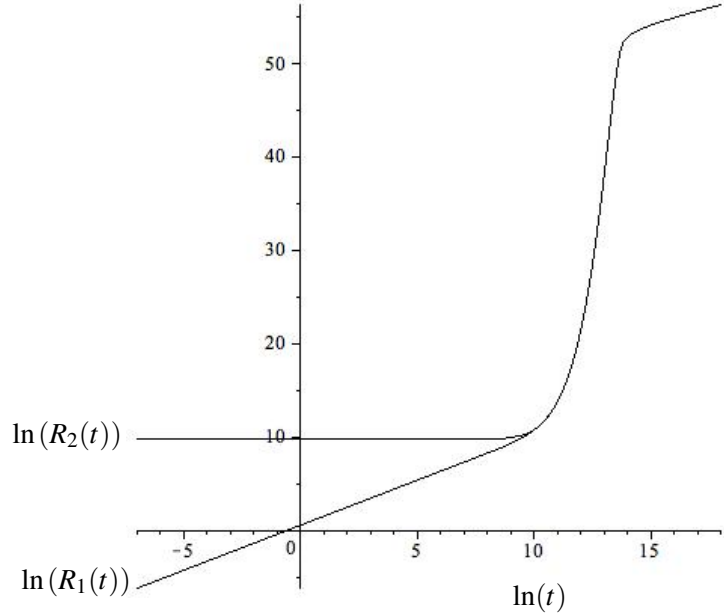
\begin{figure}
\tikzstyle{background grid}=[draw, black!50,step=.5cm]
\begin{tikzpicture}
\node (img) [inner sep=0pt,above right]
{\includegraphics[width=8cm]{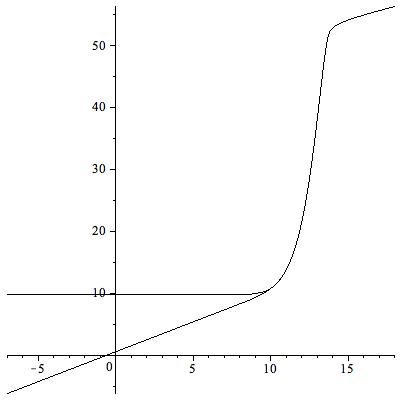}};
\draw (0:5.5cm) node {$\ln(t)$};
\draw (110:2.3cm) node {$\ln\left(R_2(t)\right)$};
\draw (162:0.7cm) node {$ \ln\left(R_1(t)\right)$};
\end{tikzpicture}
\caption[dummy1]{Dynamics of the DLH biaxial Bianchi IX model:
 evolution of the logarithm of the scale factors $R_1$ and $R_2$ in Planck
lengths $l_p$ versus log time ($t$ in units of Planck time $t_p$).}
\label{figDLH}
\end{figure}

Fig. \ref{figDLH} shows the evolution of the scale factors $R_1$ and
$R_2$ (for $t > 0$) for the viable cosmological solution presented in
\cite{DechantHobsonLasenby2009BianchiIXScalarField}, which is defined
by the initial parameters $a_0=1.2$, $b_0=18000$ and $f_0=13$ (set by
imposing a boundary condition at temporal infinity on the total
elapsed conformal time, as suggested by
\cite{LasenbyDoran2005ClosedUniversesdSandInflation}), together with
$\kappa=1$ and $m=1/64000$ (set in order to fix the normalisation of
the resulting perturbation spectrum) and $\Lambda$ set to
zero, as it is dynamically unimportant at the early time scales that we are interested in. 
This cosmological model was  obtained for a representative set of parameter
values of order unity, rather than having been fixed in order to get best
agreement with current data. These natural values were then scaled to the
seemingly less natural values given above, using a scaling property  discussed in the next paragraph. 
 In order to fix the normalisation of the perturbation
spectrum, the mass of the scalar field has to be rescaled and $b_0$
changes to a less natural value accordingly.  This choice for the mass
of the scalar field needs to be put in by hand for every current inflation model so does
not constitute any unusual fine-tuning.

As noted in \cite{DechantHobsonLasenby2009BianchiIXScalarField},
given a solution to the equations
(\ref{special1})-(\ref{specialEoM}), a family of solutions is
generated by scaling with a constant $\alpha$ and defining
\begin{equation}
\bar{R_i}(t)=\frac{1}{\alpha} R_i(\alpha t),\quad 
\bar{H_i}(t)=\alpha H_i(\alpha t),\quad
\bar{\phi}(t)=\phi(\alpha t),\quad
\bar{m}=\alpha m, \quad  
\bar{\Lambda}=\alpha ^2\Lambda.
\label{scaling}
\end{equation}
This scaling property is valuable for numerical work, as a range of
situations can be covered by a single numerical
integration. Furthermore, many physically interesting quantities turn
out to be invariant under changes in scale. This scaling property does
not, however, survive quantisation, as any quantisation prescription for the scalar field 
introduces a length scale, which breaks the scale invariance. Therefore one would have to be careful
when considering vacuum fluctuations.

In \cite{DechantHobsonLasenby2009BianchiIXScalarField} we 
demonstrated that at the time of pancaking, there is an instantaneous
reduction in the number of dimensions of the homogeneous
hypersurfaces, without a geometric singularity, or any singularities
in physical quantities.  Geodesics can extend through this point,
though some of them may wind infinitely around the topologically
closed dimension. In the light of the generic BKL-analysis mentioned
above, it is interesting to note that the dynamics of our model is
very straightforward.

\section{The Taub-NUT model}\label{TaubNUT}

The Taub-NUT model
\cite{RyanShepley1975HomogeneousRelativisticCosmologies} is a biaxial
Bianchi IX vacuum solution. Traditionally, the form of the
metric chosen to describe it does not have the form (\ref{metric}),
but is instead written
\begin{equation}
ds^2=2du\omega^1-g(u) (\omega^1)^2-e^{2\zeta(u)}\left[(\omega^2)^2+(\omega^3)^2\right],
\label{lineelmttaubnut2}
\end{equation}
where, for later convenience, we denote the standard Ryan \& Shepley
time coordinate by $u$ (rather than $t$) and replace their function
$\beta(t)$ by $\zeta(u)$.

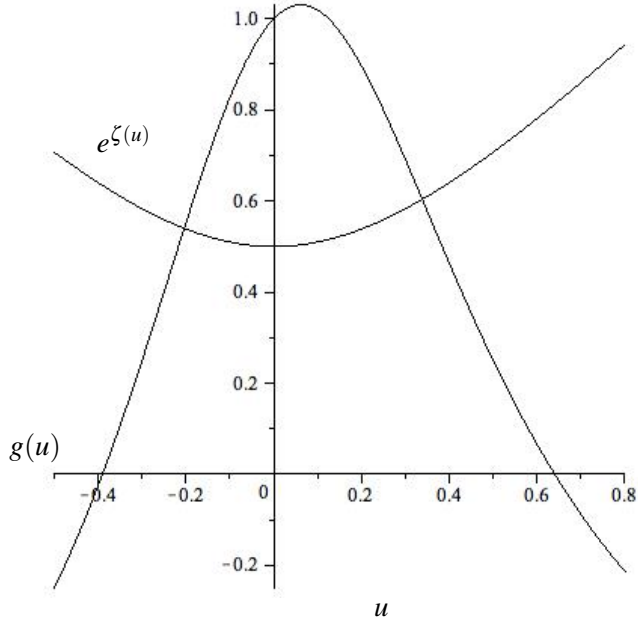
\begin{figure}
\tikzstyle{background grid}=[draw, black!50,step=.5cm]
\begin{tikzpicture}
\node (img) [inner sep=0pt,above right]
{\includegraphics[width=8cm]{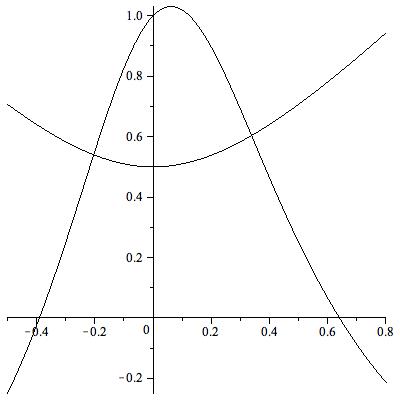}};
\draw (-2:4.5cm) node {$u$};
\draw (93:2cm) node {$g(u)$};
\draw (80:6.2cm) node {$e^{\zeta(u)}$};
\end{tikzpicture}
\caption[dummy5]{Analytic Taub-NUT solution (\ref{taubnutsol}) for $A=B=1$.}
\label{figTaubNUT}
\end{figure}

The metric (\ref{lineelmttaubnut2}) can be shown to solve the vacuum Einstein
equations provided
\begin{equation}
g(u)=\frac{Au+1-4B^2u^2}{B(4B^2u^2+1)},\qquad
e^{\zeta(u)}=\left(Bu^2+\frac{1}{4B}\right)^{\frac{1}{2}},
\label{taubnutsol}
\end{equation}
where $A$ and $B$ are arbitrary constants, which when varied lead to a
family of solutions. The above functions are plotted in
Fig. \ref{figTaubNUT} for the choice of values $A=B=1$. Note that
$e^{\zeta(u)}$ is always positive as asserted in
\cite{RyanShepley1975HomogeneousRelativisticCosmologies}, but also
that it behaves like $|u|$ for large values of $B$ or $u$, whereas
$g(u)$ has the form of an inverted parabola for small $u$ and
approaches a constant for large $u$. Essentially, $B$ measures the
smoothness of $e^{\zeta(u)}$ at the origin, whereas $A$ shifts the
centre of the approximate parabola $g(u)$. We note that there is no
analogue of the simple scaling family of solutions (\ref{scaling}) in
this setup.

The usual interpretation of Fig. \ref{figTaubNUT} is that Taub-NUT has two NUT
regions, corresponding to negative values of $g$ and one Taub region,
where $g$ is positive.  One infers that the Taub-NUT solution can be
represented as a disc that evolves into an ellipsoid and back into a
disc. In particular, it is considered to evolve from timelike open
sections in a NUT region, via lightlike sections (called Misner
bridges), to spacelike closed sections in the Taub region, back into
timelike open sections in the other NUT region. This
open-to-closed-to-open transition is not mathematically singular, but
it is incomplete, as geodesics spiral infinitely many times around the
topologically closed spatial dimension as they approach the boundary
\cite{KonkowskiHelliwellShepley1985quasiregularI,
  KonkowskiHelliwell1985quasiregularII,Carter2006Thesis, Carter1970Commutation,
MisnerTaub1969SingularityFreeEmpty, HawkingEllis1973LargeScaleStructureOfSpaceTime,
Siklos1978Whimper} (see, for instance, 
\cite{Laemmerzahl2010TaubNUTgeodesics} for a recent treatment). This type of singularity is
called `quasiregular' in the Ellis and Schmidt classification
\cite{EllisSchmidt1977SingularSpacetimes, 
Krasnikov2009QuasiregularSingularitiesTakenSeriously} {(these include the
  well-known `conical' singularities
  \cite{KonkowskiHelliwell2005Classification})}, as opposed to a
(scalar or non-scalar) curvature singularity.  The Taub-NUT solution
therefore shows the same feature of dimensional reduction as the DLH
model and similarly does not have a geometric singularity during this
collapse. Note, however, that the homogeneous hypersurfaces are only
spacelike in the Taub region, leading to problems with Taub-NUT as a
description of a homogeneous universe as one approaches the boundaries
of the Taub region (the Misner bridges).

\section{Relationship between DLH and Taub-NUT models\label{transformation}}

The similarities between the DLH and Taub-NUT models are sufficiently
striking that the connection between the two models warrants further
consideration. In particular, it is known that Taub-NUT is the only
Petrov type D homogeneous closed vacuum spacetime 
and that all Petrov D solutions are known
\cite{Stephani2003ExactSolutions}.  It is a straightforward, if
tedious, calculation to show that the DLH metric (\ref{lineelmtDLH2}) is also of
Petrov type D.  Moreover, one finds that the DLH and Taub-NUT metrics
have the same degeneracy structure in their principal curvatures,
i.e. the eigenvalues of the Riemann tensor (rather than the Weyl
tensor used in the Petrov classification), in that there are two
degenerate pairs and two singlets amongst the six real eigenvalues, as further explained
in appendix \ref{app2}.
Furthermore, the most general biaxial Bianchi IX metric is also known
to be the Pleba\'nski-Demia\'nski metric
\cite{PlebanskiDemianski1976RotatingChargedandUniformlyAccelerating}.
The similar behaviour of geodesics in both models is also suggestive. 
An analysis of geodesics in both models is contained in appendix \ref{app3}, and
more details can be found in \cite{DechantHobsonLasenby2009BianchiIXScalarField}. 
 These similarities suggest that the DLH and Taub-NUT metrics might describe
the same spacetime geometry.  In fact, as we now demonstrate, the two
metrics are indeed related by a coordinate transformation.

In general, the diffeomorphism invariance of general relativity means
it can be difficult to determine if two metrics describe genuinely
different spacetime geometries or are the same up to a diffeomorphism.
Nonetheless, in the latter case, such a diffeomorphism can be found
from considering several scalar invariants. In general, four
independent scalar invariants allow us to fix the diffeomorphism
between the two spacetimes. A further invariant can then be used to
check consistency or to derive a contradiction. In our present case,
both models are homogeneous and so scalar invariants are functions of
time only. Thus, by considering how a single curvature invariant,
e.g. the Ricci scalar, transforms, one can straightforwardly identify
an appropriate coordinate transformation linking the two metrics,
which amounts simply to a time rescaling.

Starting with the DLH metric (\ref{lineelmtDLH2}), let us consider the time
rescaling transformation 
\begin{equation}
u=\int_0^t \frac{1}{2}R_1(t')dt' \equiv f^{-1}(t)
\label{timerep2}
\end{equation}
and define the functions $P_i(u) \equiv R_i(f(u)) = R_i(t)$. 
The lower limit on the integral is chosen for convenience such that 
$u=0$ when $t=0$. One could make another choice, but this simply adds
a constant shift to the value of $u$. Moreover,
let us shift the non-degenerate spatial one-form by a timelike part to
obtain
\begin{equation}
\sigma^1\equiv\omega^1+\frac{2}{R_1(t)}dt, \quad
\sigma^2\equiv\omega^2, \quad \sigma^3\equiv\omega^3.
\label{oneformshift2}
\end{equation}
The shift is necessary to absorb unwanted terms arising from the time
reparameterisation into a redefinition of the one-forms. As long as
the one-forms are purely spatial to start with, the timelike component
that we have added to the first one-form commutes through, such that
this redefined set of one-forms still obeys the $\mathfrak{su}(2)$ commutation
relations and is therefore a valid set to describe homogeneous
hypersurfaces. One thus obtains the metric
\begin{equation}
ds^2=2du\sigma^1-{\textstyle\frac{1}{4}} P_1(u)^2
(\sigma^1)^2-{\textstyle\frac{1}{4}}P_2(u)^2\left[(\sigma^2)^2+(\sigma^3)^2\right].
\label{lineelmttaubnut3}
\end{equation}
Comparing this expression with the Taub-NUT metric
(\ref{lineelmttaubnut2}), one sees that they have the same form, but
with $\frac{1}{4}P_1^2$ replacing $g$ and $\frac{1}{4}P_2^2$ replacing
$e^{2\zeta}$.  It is straightforward to verify that the time rescaling
(\ref{timerep2}) maps all curvature scalars for the DLH metric, such
as the Ricci scalar (by construction), the Euler-Gauss-Bonnet
invariant
\begin{equation}
R_{GB}^2=R^2_{\mu\nu\sigma\tau}-4R^2_{\mu\nu}+R^2,
\label{GB}
\end{equation}
the Chern-Pontryagin scalar
\begin{equation}
K_{CP}={{}^\star R}_{\mu\nu\sigma\tau} R^{\mu\nu\sigma\tau},
\label{CP}
\end{equation}
and the eigenvalues of the Riemann and Weyl tensors (of course they
are not necessarily independent) into those obtained for the Taub-NUT
metric (c.f.  appendix \ref{app2}). Moreover, we note that the form of the resulting Einstein field
equations do not depend on the concrete realisation of the $\mathfrak{su}(2)$
1-forms used in the metric; one simply requires that they obey the
$\mathfrak{su}(2)$ algebra commutation relations, which we have guaranteed by
construction. A mapping between the geodesic equations in both models, for  
the explicit realisation used in \cite{DechantHobsonLasenby2009BianchiIXScalarField}, 
is exhibited in appendix \ref{app3}.

The time rescaling transformation (\ref{timerep2}) is valid
independently of any concrete choice for the metric functions.
Nonetheless, it can be seen that whenever $R_1(t)$ (or $P_1(u)$) goes
through zero, as is the case at pancaking events of the DLH model (or
on the Misner bridges in the Taub-NUT model), this coordinate
transformation will be problematic.  The transformation itself is not
singular, but one sees that when the integrand $R_1$ changes sign
(such as at the pancaking events), the definition of $u$ becomes
multivalued: the value of the integral for $u$ will begin to decrease
as $t$ increases. Thus there is only a one-to-one correspondence
between $u$ and $t$ as long as $R_1$ and $P_1$ do not change
sign. Furthermore, the definition of the 1-forms (\ref{oneformshift2})
goes singular at the pancaking events also indicating a problem with
$u$ as a measure of time (c.f. Section \ref{TNUTSF}).

We also note that from (\ref{timerep2}) we have that
\begin{equation}
\frac{du}{dt}=\frac{R_1(t)}{2}=\frac{P_1(u)}{2},
\label{dudt}
\end{equation}
and thus the inverse transformation is  simply given by
\begin{equation}
t=\int_0^u \frac{2}{P_1(u')}du';
\label{timerep3}
\end{equation}
Hence DLH time $t$ is essentially conformal time for Taub-NUT. This
seems rather odd, as DLH is the natural generalisation of closed FRW
models, and one is usually interested in conformal time associated
with those cosmologically interesting solutions, which would make the
usual cosmological conformal time doubly conformal Taub-NUT time.

\section{The reparameterised Taub-NUT model}\label{repTNUT}

The new form (\ref{lineelmttaubnut3}) for the Taub-NUT metric differs
significantly from the traditional form (\ref{lineelmttaubnut2}).  We
believe that our new form in terms of the scale factors is physically
more meaningful, as the squares of the scale factors premultiply the
invariant one-forms, as in the DLH metric, or in the FRW special case,
to give physical distances on the homogeneous spacelike slices.

The differences between our form (\ref{lineelmttaubnut3}) and the
traditional form (\ref{lineelmttaubnut2}) brings into question the
usual interpretation of the Taub-NUT vacuum solution
(\ref{taubnutsol}).  In particular, we see that the traditional $g$
function is rather unnatural, as it corresponds to the square of a
scale factor, rather than the more physically meaningful scale factor
itself.  Moreover, as the square of a real number, the function $g$
should always be positive. Hence we should not allow the $g$ function
to go negative, and must instead select the positive branch at all
times. This yields a very different picture from the alleged
open-to-closed-to-open transition. If one simply took the modulus of
the solution in Fig. \ref{figTaubNUT}, such that the $g$-function is
positive throughout, the resulting model is not a solution of the Einstein equations, 
unless $e^{2\zeta}$ is also allowed to flip sign and become negative,
thus raising a new problem. 
Rather, one should solve afresh the Einstein equations using 
the reparameterised Taub-NUT metric as the Ansatz. 

In our parametrisation (\ref{lineelmttaubnut3}), in terms of the scale factors
$P_i(u)$, with associated Hubble functions $K_i(u)\equiv {P}_i'/P_i$
(where a prime denotes $d/du$) and scalar field $F(u)\equiv \phi
(f(u))=\phi (t)$, the Einstein field equations yield the dynamical
equations
\begin{equation}
4P_1^2K_1'+8P_1^2K_1^2-2P_1^2K_2^2-4\kappa m^2F^2+
\kappa P_1^2F'^2 +4P_1^2K_1K_2-8\Lambda+\frac{8}{P_2^2}\left(5\frac{P_1^2}{P_2^2}
-4\right)=0\label{TNUT1}\end{equation}
and
\begin{equation}
4P_1^2K'_2+6P_1^2K_2^2-4\kappa m^2F^2+\kappa P_1^2F'^2 +4P_1^2K_1K_2-8\Lambda-\frac{8}{P_2^2}\left(3\frac{P_1^2}{P_2^2}-4\right)=0,\label{TNUT2}\end{equation}
as well as the Friedmann equation (or Hamiltonian constraint)
\begin{equation}
-2P_1^2K_2^2+4\kappa m^2F^2 +\kappa P_1^2F'^2-4P_1^2K_1K_2+8\Lambda+\frac{8}{P_2^2}\left(\frac{P_1^2}{P_2^2}-4\right)=0,\label{TNUTFr}\end{equation}
and, in general, the equation of motion for the scalar field 
\begin{equation}
 P_1^2F''+2P_1^2 (K_1+K_2)F'+4m^2F=0.
\label{TNUTEoM}
\end{equation}
It is straightforward to show that, as expected, these equations
also result from directly applying the time rescaling 
transformation (\ref{timerep2}) and associated function redefinitions to the
evolution equations (\ref{special1})-(\ref{specialEoM}) of the DLH model.

We will solve the above system of equations, with and without scalar
matter, in Section~\ref{scenarios}. For the moment, however, we
concentrate on the issue of scaling solutions.
Considering the family of solutions in the DLH model related by
(\ref{scaling}), one could use the time rescaling transformation
(\ref{timerep2}) to map these solutions into our reparameterised
Taub-NUT model, thereby constructing an analogous family of Taub-NUT
solutions. In general, however, these will not be related by a simple
scaling relation as in (\ref{scaling}).  Nonetheless, our
reparameterised version of Taub-NUT does admit directly a family of
solutions that are related by a straightforward scaling of the form
\begin{equation}
\bar{P}_i(u)=\frac{1}{\sqrt{\beta}} P_i(\beta u),\quad
\bar{K}_i(u)=\beta K_i(\beta u),\quad
\bar{F}(u)=F(\beta u),\quad
\bar{m}=\sqrt{\beta} m, \quad 
\bar{\Lambda}=\beta \Lambda.
\label{TNUTscaling}
\end{equation}
Comparing this result with the corresponding scaling invariance
(\ref{scaling}) of solutions in the DLH model, we see that they both
arise from a simple constant time rescaling of the form $t \to
\alpha\bar{t}$ (in the DLH model) or $u \to \beta \bar{u}$ (in the
reparameterised Taub-NUT model). The relative square root between
$\alpha$ and $\beta$ results from the fact that the Taub-NUT metric is
linear in the `time' parameter $u$ whereas the DLH metric is quadratic
in $t$.

The integral definition (\ref{timerep2}) of $u(t)$ precludes a
straightforward identification of $u(\alpha t)$ with $\alpha
u(t)$. One can, however, extend the diffeomorphism (\ref{timerep2})
to take us from scaled DLH to scaled Taub-NUT 
by writing
\begin{equation}
\beta\bar{u}=\int_0^{\alpha\bar{t}} \frac{1}{2}R_1(t')dt'
\label{timereptausigma}
\end{equation}
and 
\begin{equation}
\sigma^1\equiv\omega^1+\frac{2\alpha}{R_1(\alpha\bar{t})}d\bar{t},\quad 
\sigma^2\equiv\omega^2, \quad 
\sigma^3\equiv\omega^3,
\label{oneformshifttausigma}
\end{equation}
and defining $P_i(\beta\bar{u}) \equiv R_i(\alpha\bar{t})$ and 
$F(\beta\bar{u})=\phi (\alpha\bar{t})$.

\section{Comparison of DLH and reparamaterised Taub-NUT models
\label{scenarios}}

Although there exists a coordinate transformation linking the DLH and
reparameterised Taub-NUT metrics, this transformation is multivalued
when pancaking events occur in the DLH model and at the Misner bridges
in the Taub-NUT model; this leads to differences in the physical
interpretation of the corresponding cosmological solutions.  In this
section, we therefore consider, in turn, the four cases of vacuum and
scalar field matter solutions in both DLH and reparameterised
Taub-NUT.  We contend that the DLH set-up is the more physically
meaningful.  Anticipating this conclusion, we start by considering the
DLH vacuum model as the most fundamental setup.

\subsection{DLH vacuum solution \label{DLHvac}}

Setting the scalar field $\phi=0$ in the DLH evolution equations
(\ref{special1})-(\ref{specialEoM}), and also assuming $\Lambda=0$ for
simplicity, as it is unimportant dynamically at early times, yields the system
\begin{equation}
2\dot{H}_{2}+3H_{2}^{2}=\frac{1}{R_{2}^{2}}\left(3\frac{R_{1}^{2}}{R_{2}^{2}}-4\right),\label{vacspecial1}\end{equation}
\begin{equation}
2H_{1}^{2}+2\dot{H}_{1}-H_{2}^{2}+2H_{1}H_{2}=-\frac{1}{R_{2}^{2}}\left(5\frac{R_{1}^{2}}{R_{2}^{2}}-4\right),\label{vacspecial2}\end{equation}
\begin{equation}
H_{2}^{2}+2H_{1}H_{2}=\frac{1}{R_{2}^{2}}
\left(\frac{R_{1}^{2}}{R_{2}^{2}}-4\right),\label{vacspecialFr}\end{equation}
which can be solved analytically, albeit in terms of elliptic
integrals. The detailed derivation of the analytic form is given in 
 appendix \ref{app1}. In short, the Einstein equations can be integrated to
find one scale factor $R_1$ and time $t$ in terms of the other scale
factor $R_2$ as
\begin{equation}
{R_1(t)=\pm\frac{R_2(t)\dot{R}_2(t)}{\sqrt{\mu R_2(t)^2-1}}},
\label{R1itoR2b}
\end{equation}
and
\begin{equation}
t=\sqrt{\frac{i}{2a_0}\frac{x^2+a_0^2}{a_0^2+4}}y
+\sqrt{\frac{a_0}{8i}} 				E\left(y;i\right)
+\sqrt{\frac{ia_0}{32}}(a_0+2i)		F\left(y; i\right) 
-\sqrt{\frac{i}{32a_0}}(a_0^2-4)	\Pi\left(y; \frac{2}{ia_0};i\right),
\label{ellintsol2}
\end{equation}
where $E(z; k)$, $F(z; k)$ and $\Pi(z;\nu;k)$ are Legendre's three
normal forms, $\mu$ an integration constant that can be fixed in terms
of the initial conditions, and $x$ and $y$ are functions of $R_2$
defined in  appendix \ref{app1}.  Alternatively, the system of evolution
equations can be solved numerically; comparison of the numerical and
analytical solutions shows very good agreement (see
Fig. \ref{figDLHvacperanalnum}).

\begin{figure}
\tikzstyle{background grid}=[draw, black!50,step=.5cm]
\begin{tikzpicture}
\node (img) [inner sep=0pt,above right]
{\includegraphics[width=8cm]{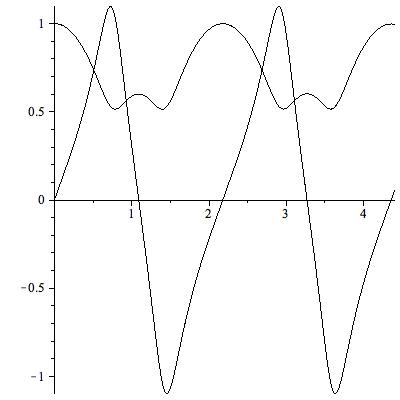}};
\draw (-2:4.5cm) node {$t$};
\draw (95:4cm) node {$R_1$};
\draw (93:7.6cm) node {$R_2$};
\end{tikzpicture}
\caption[dummy5]{Periodic `DLH vacuum' solution for the values $a_0=1$ and $b_0=1$. }
\label{figDLHvacperiodic2}
\end{figure}
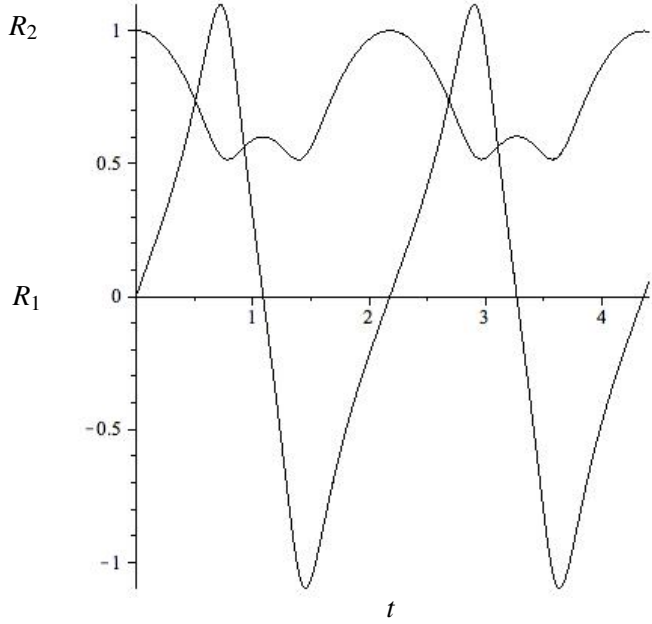

The solution is periodic in $t$, as shown in
Fig. \ref{figDLHvacperiodic2}, which is a numerical solution
with  boundary conditions analogous to the pancaking solution 
in the case with the scalar field (\ref{seriesansatz}), i.e. 
\begin{equation}
R_1(0)=0, \,\,\, \dot{R}_1(0)=a_0,\,\,\, R_2(0)=b_0,\,\,\,\dot{R}_2(0)=0,
\label{ellBCs}
\end{equation}
and we make the convenient choice $a_0=1$, $b_0=1$.
In addition to the periodicity, the solution is symmetric about the
two pancaking points in each period, with a parity inversion in $R_1$
(which is linear near pancaking events) and $R_2$ being even, in
agreement with our previous pancaking DLH solution in
\cite{DechantHobsonLasenby2009BianchiIXScalarField}. This is an
interesting `cyclic' model that repeats indefinitely. As we will see
in the following section, it is stable to inclusion of a perturbing
scalar field, but as the scalar field becomes heavier and/or denser,
strict periodicity is broken.

\subsection{DLH solution with a scalar field}\label{DLH_SF}

This physical set-up is, of course, that which we originally
considered in \cite{DechantHobsonLasenby2009BianchiIXScalarField}.  As
re-iterated in Section~\ref{DLH}, for some (quite natural) assumed
values of the initial conditions and scalar field mass, such a model
leads to a viable cosmology.  It is of interest here, however, to
investigate the transition from the cyclic DLH vacuum solution
outlined above to the viable cosmological model by `gradually'
introducing the scalar field.  This can be achieved by allowing the
scalar field to become progressively denser or the mass of the scalar
field heavier. Here the same boundary conditions are assumed as
previously for the pancaking series solution (\ref{seriesansatz}),
i.e. oddness for $R_1$ and evenness for $R_2$ and $\phi$.

\begin{figure}
       \begin{tabular}{@{}c@{ }c@{ }c@{ }}
		\begin{tikzpicture}
		\node (img) [inner sep=0pt,above right]
		{\includegraphics[height=0.25\linewidth]{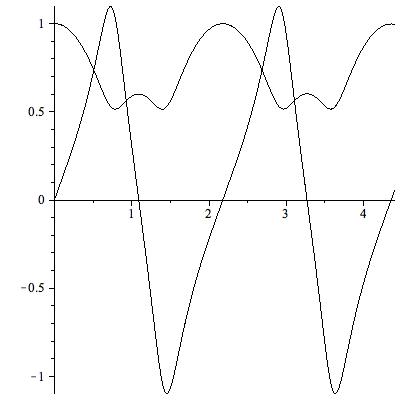}};
		\draw (-2:2.5cm) node {$t$};
		\draw (95:2.2cm) node {$R_1$};
		\draw (93:3.5cm) node {$R_2$};
		\end{tikzpicture}&
 	\begin{tikzpicture}
	\node (img) [inner sep=0pt,above right]
	{\includegraphics[height=0.25\linewidth]{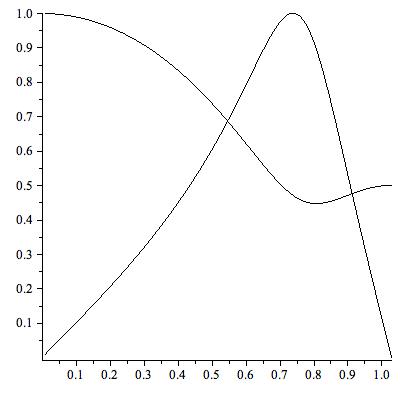}};
	\draw (-2:2.5cm) node {$t$};
	\draw (95:0.5cm) node {$R_1$};
	\draw (93:3.5cm) node {$R_2$};
	\end{tikzpicture}  & 
   	\begin{tikzpicture}
	\node (img) [inner sep=0pt,above right]
	{\includegraphics[height=0.25\linewidth]{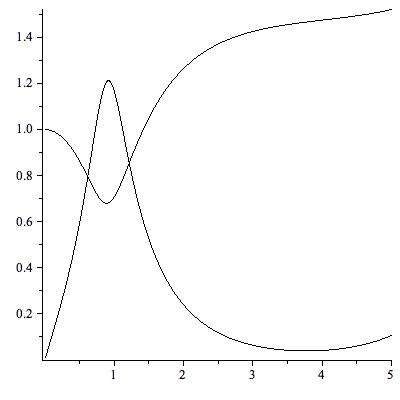}};
	\draw (-2:2.5cm) node {$t$};
	\draw (95:0.5cm) node {$R_1$};
	\draw (93:2.5cm) node {$R_2$};
	\end{tikzpicture}\\
     \tiny a) & \tiny b)  & \tiny c) \\
		\begin{tikzpicture}
		\node (img) [inner sep=0pt,above right]
		{\includegraphics[height=0.25\linewidth]{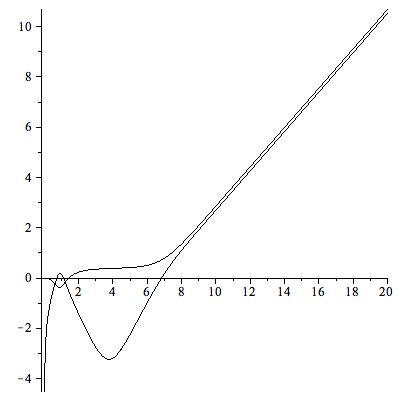}};
		\draw (-2:2.5cm) node {$t$};
		\draw (95:0.4cm) node {$\ln R_1$};
		\draw (93:1.0cm) node {$\ln R_2$};
		\end{tikzpicture}&
 	\begin{tikzpicture}
	\node (img) [inner sep=0pt,above right]
	{\includegraphics[height=0.25\linewidth]{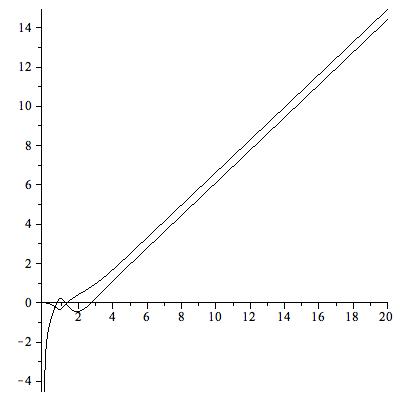}};
	\draw (-2:2.5cm) node {$t$};
	\draw (95:0.4cm) node {$\ln R_1$};
	\draw (93:1.0cm) node {$\ln R_2$};
	\end{tikzpicture}  & 
   	\begin{tikzpicture}
	\node (img) [inner sep=0pt,above right]
	{\includegraphics[height=0.25\linewidth]{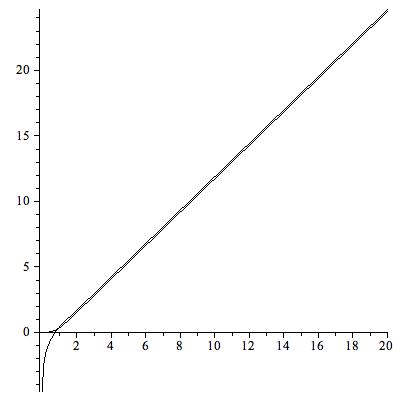}};
	\draw (-2:2.5cm) node {$t$};
	\draw (95:0.4cm) node {$\ln R_1$};
	\draw (93:0.7cm) node {$\ln R_2$};
	\end{tikzpicture}\\
     \tiny d) & \tiny e)  & \tiny f) \\
  \end{tabular}
  \caption[dummy5]{Gradually introducing a light scalar field by
    increasing $\phi_0$, whilst
    keeping the other parameters fixed at $\kappa=1$, $\Lambda=0$,
    $m=1/64000$, $a_0=1$, $b_0=1$: a) shows the
    vacuum model $\phi_0=0$; b) $\phi_0=4$; c) and d) both have
    $\phi_0=1.22\times 10^5$; e) 
    $\phi_0=1.3\times 10^5$ and (f) $\phi_0=2\times 10^5$.}
\label{figSFinDLH}
\end{figure}
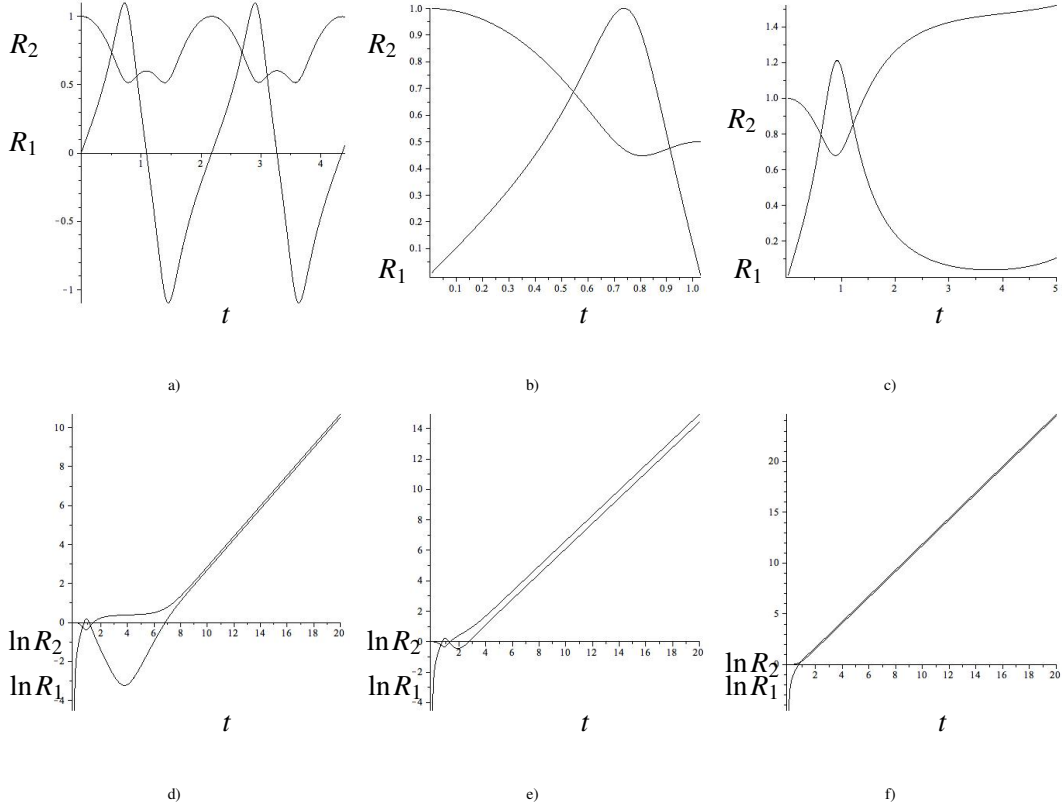

In order to assess the effect of increasing scalar field energy
density, the remaining parameters are kept constant at $\kappa=1$,
$\Lambda=0$, $m=1/64000$, $a_0=1$, $b_0=1$, whilst $\phi_0$ is varied
over the range $0\le \phi_0\le 2\times 10^5$.  Fig. \ref{figSFinDLH}
shows the vacuum solution in panel (a), and a small perturbation
thereof, with $\phi_0=4$, in panel (b). As the scalar field is
increased, $R_1$ turns around (panel c) and inflation is produced 
(panel d) (both
for $\phi_0=1.22\times 10^5$). Note that, from panel (d) onwards, we
increase the range in $t$ and take logarithms of the scale factors, to
account for the fact that inflation is produced. Panels (e) and (f)
show how higher initial scalar field energy densities produce more
inflation (for $\phi_0=1.3\times 10^5$ and $\phi_0=2.0\times 10^5$
respectively).

\begin{figure}
       \begin{tabular}{@{}c@{ }c@{ }c@{ }}
		\begin{tikzpicture}
		\node (img) [inner sep=0pt,above right]
		{\includegraphics[height=0.25\linewidth]{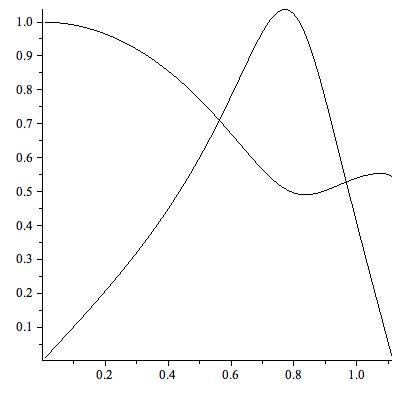}};
		\draw (-2:2.5cm) node {$t$};
		\draw (95:0.5cm) node {$R_1$};
		\draw (93:3.8cm) node {$R_2$};
		\end{tikzpicture}&
 	\begin{tikzpicture}
	\node (img) [inner sep=0pt,above right]
	{\includegraphics[height=0.25\linewidth]{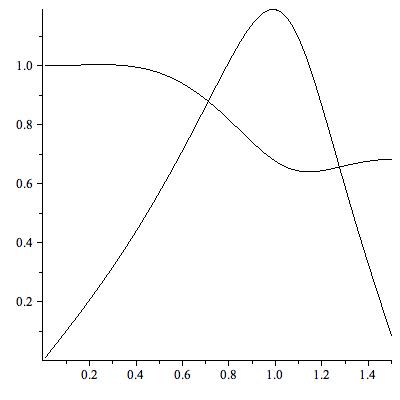}};
	\draw (-2:2.5cm) node {$t$};
	\draw (95:2.5cm) node {$R_1$};
	\draw (93:3.6cm) node {$R_2$};
	\end{tikzpicture}  & 
   	\begin{tikzpicture}
	\node (img) [inner sep=0pt,above right]
	{\includegraphics[height=0.25\linewidth]{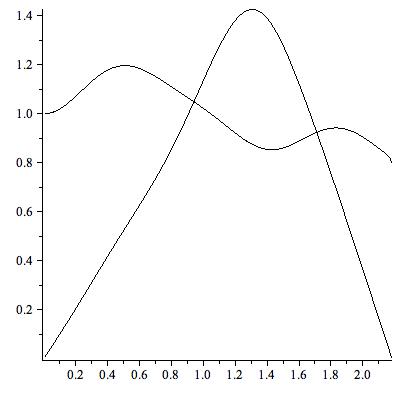}};
	\draw (-2:2.5cm) node {$t$};
	\draw (95:2.4cm) node {$R_1$};
	\draw (93:3.4cm) node {$R_2$};
	\end{tikzpicture}\\
     \tiny a) & \tiny b)  & \tiny c) \\
		\begin{tikzpicture}
		\node (img) [inner sep=0pt,above right]
		{\includegraphics[height=0.25\linewidth]{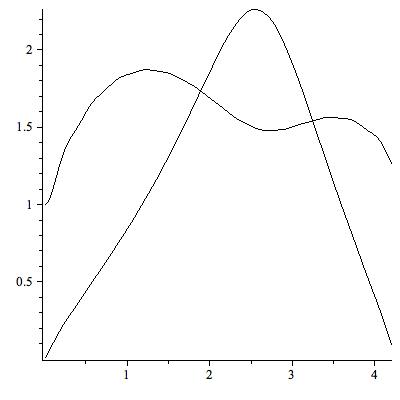}};
		\draw (-2:2.5cm) node {$t$};
		\draw (95:1.0cm) node {$R_1$};
		\draw (93:2.4cm) node {$R_2$};
		\end{tikzpicture}&
 	\begin{tikzpicture}
	\node (img) [inner sep=0pt,above right]
	{\includegraphics[height=0.25\linewidth]{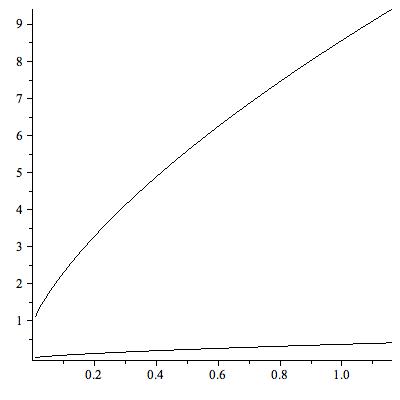}};
	\draw (-2:2.5cm) node {$t$};
	\draw (95:0.4cm) node {$R_1$};
	\draw (93:1.0cm) node {$R_2$};
	\end{tikzpicture}  & 
   	\begin{tikzpicture}
	\node (img) [inner sep=0pt,above right]
	{\includegraphics[height=0.25\linewidth]{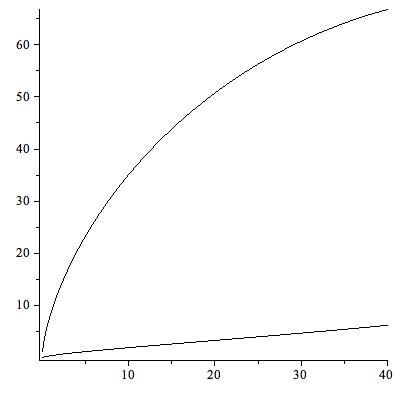}};
	\draw (-2:2.5cm) node {$t$};
	\draw (95:0.2cm) node {$R_1$};
	\draw (93:0.7cm) node {$R_2$};
	\end{tikzpicture}\\
     \tiny d) & \tiny e)  & \tiny f) \\
  \end{tabular}
  \caption[dummy5]{Gradually increasing the mass of the scalar field
    whilst keeping the other parameters fixed at $\kappa=1$,
    $\Lambda=0$, $\phi_0=1$, $a_0=1$, $b_0=1$: (a) shows the model for
    $m=1$; (b) $m=3$; (c) $m=5$; (d) $m=10$; (e) and (f) are both for
    $m=100$, with the latter covering a wider range in $t$.}
\label{figheavySFinDLH}
\end{figure}

 Similarly, in order to study the effects of increasing the mass of the scalar
 field (Fig. \ref{figheavySFinDLH}) (rather than its density), the other parameters are kept
 fixed at $\kappa=1$, $\Lambda=0$, $\phi_0=1.0$, $a_0=1$, $b_0=1$, 
whilst $m$ takes the values: (a) $m=1$, (b)
 $m=3$, (c) $m=5$, (d) $m=10$ and (e) $m=100$. Panel (f) shows
the $m=100$ case for a wider range in $t$.

For a very light or diffuse scalar field, the equation of motion
\begin{equation}
m^{2}\phi+\left(H_{1}+2H_{2}\right)\dot{\phi}+\ddot{\phi}=0
\label{generalscalarEoM}
\end{equation}
is approximately satisfied by a constant scalar field, as the mass
term is suppressed. This essentially eliminates the scalar field from
the problem, so that we recover the vacuum case with its periodic
behaviour.  As the scalar field gets denser (see
Fig. \ref{figSFinDLH}) or the mass of the scalar field heavier (see
Fig. \ref{figheavySFinDLH}) (but still subject to the same boundary conditions), 
the deviations from the vacuum case
grow, and eventually can no longer be considered small. The behaviour
is then no longer periodic, and smoothly changes qualitatively,
eventually yielding the inflationary cosmological solution described
in \cite{DechantHobsonLasenby2009BianchiIXScalarField}.

\subsection{Taub-NUT vacuum solution}

The usual derivation of the Taub-NUT vacuum solution is obtained by
substituting the standard form of the metric (\ref{lineelmttaubnut2})
into the Einstein equations to yield the family of solutions
(\ref{taubnutsol}). We now have two further methods for constructing
this solution: by mapping the DLH vacuum solution using the
diffeomorphism (\ref{timerep2}); and by substituting our
reparameterised form (\ref{lineelmttaubnut3}) of the Taub-NUT metric
into the Einstein equations and solving the resulting system of
equations given in Section~\ref{repTNUT}. These two alternatives
approaches offer different physical insights into the nature of the
Taub-NUT vacuum solution, and are considered below.

\subsubsection{Mapping the DLH vacuum solution using the diffeomorphism} \label{MapDLH}

Conceptually, the simplest way to arrive at the vacuum solution for
the reparameterised Taub-NUT metric is to apply the diffeomorphism
(\ref{timerep2}) to the DLH vacuum solution discussed in
Section~\ref{DLHvac}.  In practice, however, this rather complicated,
as it involves elliptic integrals. For the sake of simplicity, we
therefore concentrate on the period near (one of) the pancaking
events, which will be sufficient to unearth some interesting
properties of the mapping.

In the vicinity of a pancaking event, $R_1$ has the simple linear
behaviour
\begin{equation}
R_1(t)=a_0t.
\label{pancakeR1}
\end{equation}
Hence, the corresponding coefficient in the DLH metric
(\ref{lineelmtDLH2}), which is proportional to $R_1^2 \propto t^2$, is
a smooth function that touches zero at the pancaking (see
Fig. \ref{figDLHpancR1}).
\begin{figure}
\begin{tabular}{@{}c@{ }}
	\tikzstyle{background grid}=[draw, black!50,step=.5cm]
	\begin{tikzpicture}
	\node (img) [inner sep=0pt,above right]
	{\includegraphics[width=6cm]{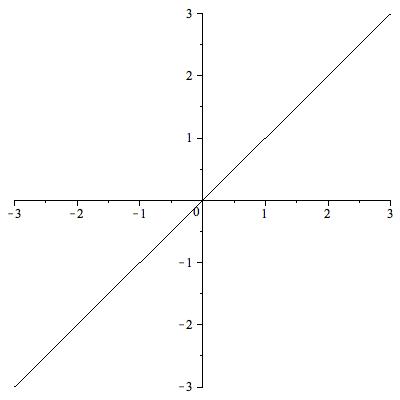}};
	\draw (-2:4.5cm) node {$t$};
	\draw (100:5cm) node {$R_1$};
	\end{tikzpicture}
    \tikzstyle{background grid}=[draw, black!50,step=.5cm]
	\begin{tikzpicture}
	\node (img) [inner sep=0pt,above right]
	{\includegraphics[width=6cm]{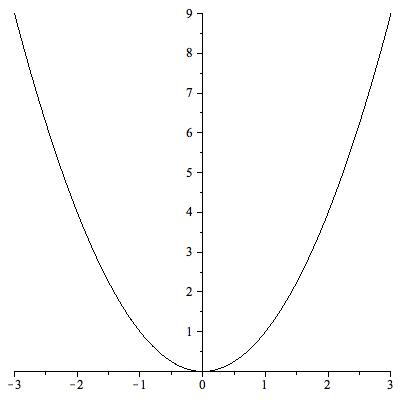}};
	\draw (-2:4.5cm) node {$t$};
	\draw (100:5cm) node {$R_1^2$};
	\end{tikzpicture}
  \end{tabular}
\caption[dummy5]{Near-pancake limit of the DLH solution: $R_1(t)$ and $R_1^2(t)$}
\label{figDLHpancR1}
\end{figure}
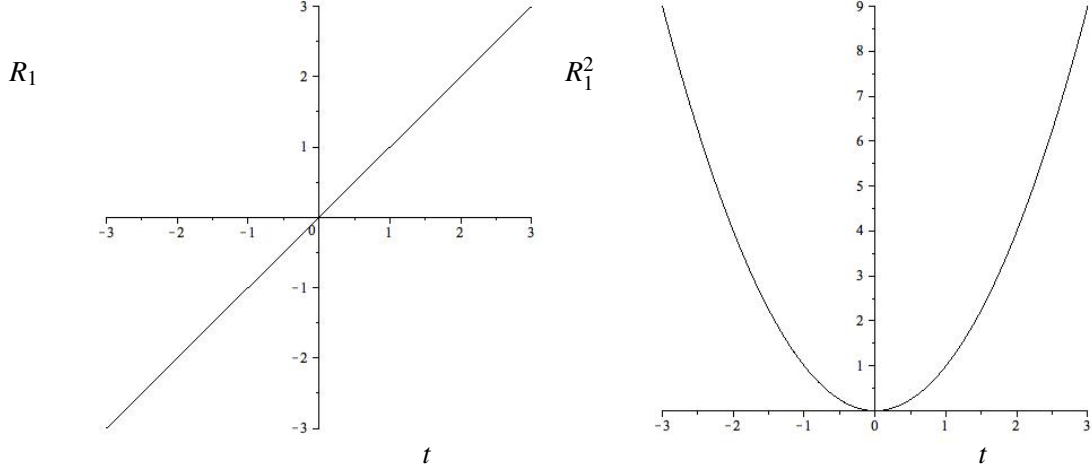
\begin{figure}
\begin{tabular}{@{}c@{ }}
	\tikzstyle{background grid}=[draw, black!50,step=.5cm]
	\begin{tikzpicture}
	\node (img) [inner sep=0pt,above right]
	{\includegraphics[width=6cm]{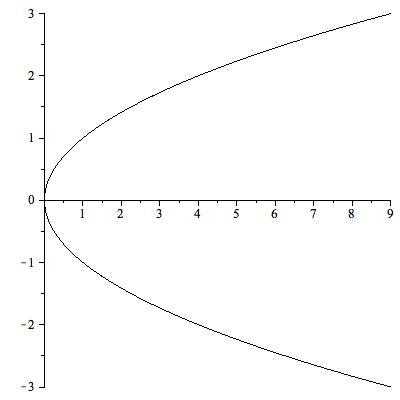}};
	\draw (-2:4.5cm) node {$u=\frac{a_0}{4}t^2$};
	\draw (100:5cm) node {$P_1$};
	\end{tikzpicture}
    \tikzstyle{background grid}=[draw, black!50,step=.5cm]
	\begin{tikzpicture}
	\node (img) [inner sep=0pt,above right]
	{\includegraphics[width=6cm]{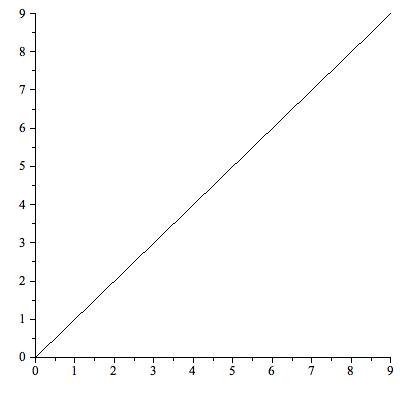}};
	\draw (-2:4.5cm) node {$u=\frac{a_0}{4}t^2$};
	\draw (100:5cm) node {$P_1^2$};
	\end{tikzpicture}
  \end{tabular}
\caption[dummy5]{Near-pancake limit of the Taub-NUT solution: $P_1(u)$ and $P_1^2(u)$}
\label{figTNUTpancP1}
\end{figure}
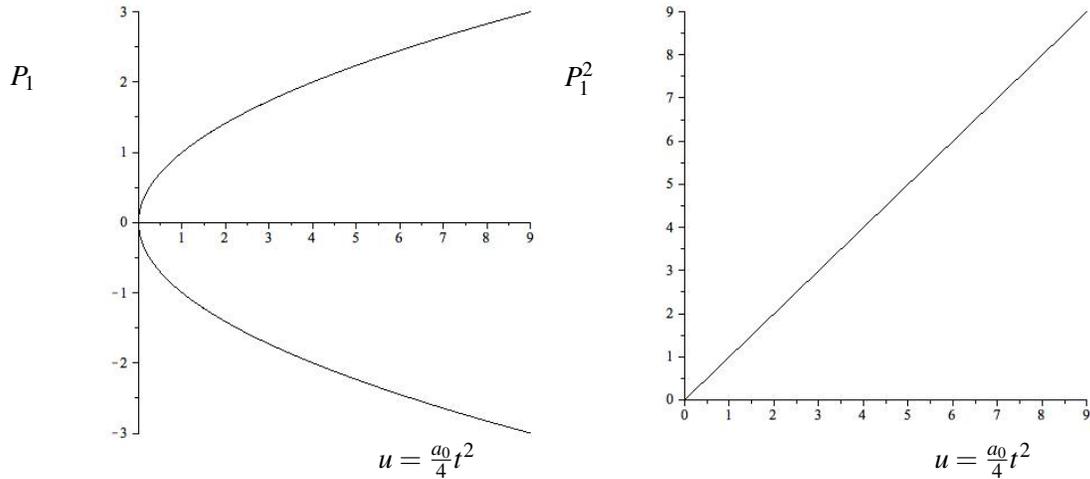
To map this solution to the reparameterised Taub-NUT model,
we first use the definition of $u$ in the diffeomorphism
(\ref{timerep2}) to obtain
\begin{equation}
u=\int_0^t \frac{1}{2}a_0 t'dt'=\frac{1}{4}a_0t^2,
\label{pancakeu}
\end{equation}
in this limit. Thus $u$ is always positive
(assuming $a_0>0$), whilst $t$ takes
both positive and negative values.  Also, since $R_1(t)=P_1(u)$, this
yields
\begin{equation}
P^2_1(u)=a^2_0t^2=4{a_0}{u}.
\label{pancakeP1}
\end{equation}
We see immediately that the innocuous pancaking of the DLH solution
now appears as $\pm\sqrt{u}$ behaviour in the corresponding Taub-NUT
solution (see Fig. \ref{figTNUTpancP1}); this also matches the lowest
order term in a series expansion of a solution to the Taub-NUT
Einstein field equations (\ref{TNUT1})-(\ref{TNUTEoM}). From the
multivaluedness, we see that $u$ is ill-defined as a time variable, as
we cannot access negative values of $u$ here, whereas $P_1(u)$ is
multivalued as a function of $u$.  Even more pathologically, $P_1^2$
(which essentially plays the same role as $g(u)$) seems to originate
only at $u=0$, and then extend out to infinity in a straight line.  In
terms of the description in terms of time $t$, $P_1^2$ comes in from
infinity, reaches the origin, and then traces back on itself.  This
already hints at an observation that will be made more precise below:
our inability to access negative values of $u$ presumably amounts to
those values corresponding to some Euclidean, imaginary time
coordinate obtained by a Wick rotation from the physical time $t$.  Of
course, the behaviour described here depends on the exact choice of
the lower limit of the integration in the definition of $u$, but
making a different choice does not avert the problem; it simply adds a 
constant offset to $u$.

\subsubsection{Direct solution of Einstein equations for
reparameterised Taub-NUT metric}

Setting the scalar field $F=0$ (and $\Lambda=0$) in
(\ref{TNUT1})-(\ref{TNUTFr}), we can, in complete analogy with Section
\ref{DLHvac}, solve directly for the Taub-NUT model parameterised in
terms of $P_1$ and $P_2$.  Note that we have already shown that for
this coordinatisation there exists a simple scaling solution
(\ref{TNUTscaling}), contrary to the original Ryan-Shepley form
(\ref{taubnutsol}).

The vacuum equations 
\begin{equation}
4K_1'+8K_1^2-2K_2^2+4K_1K_2+\frac{8}{P_1^2P_2^2}\left(5\frac{P_1^2}{P_2^2}
-4\right)=0,\label{TNUTvac1}\end{equation}
\begin{equation}
4K'_2+6K_2^2 +4K_1K_2-\frac{8}{P_1^2P_2^2}\left(3\frac{P_1^2}{P_2^2}-4\right)=0,\label{TNUTvac2}\end{equation}
\begin{equation}
-2K_2^2-4K_1K_2+\frac{8}{P_1^2P_2^2}\left(\frac{P_1^2}{P_2^2}-4\right)=0,\label{TNUTvacFr}\end{equation}
can be immediately integrated by again taking an appropriate
combination of a dynamical equation and the Friedmann equation to give
\begin{equation}
P_2(u)=\left(\frac{1+\mu^2(u+u_0)^2}{\mu}\right)^{\frac{1}{2}},
\label{taubnutsol2}
\end{equation}
where $\mu$ is a constant of integration. Note that $P_2(u)$
never goes to zero.
In fact this form for $P_2(u)$ is nearly identical to the square-root
of the standard form for $g(u)$ in the Taub-NUT solution
given in (\ref{taubnutsol}). The slight difference occurs because our
time coordinate $u$ differs to the one used by Ryan \& Shepley, in
general, by a relative shift $u_0$.  In (\ref{taubnutsol}), $u_0$ was
chosen to be zero such that the function was symmetric around
$u=0$. Here, we will instead enforce our usual boundary conditions on
$P_1$, namely that it passes through zero when $u=0$, with a certain
slope.  Substituting the expression for $P_2$ into the Friedmann
constraint allows us to integrate to find $P_1^2$, with some
additional integration constant $\beta$. Now choosing
$u_0=-\frac{4}{\beta\mu}$ such that $P_1$ vanishes at $u=0$, we find
\begin{equation}
P_1^2(u)=\left(\frac{\beta u(-4\beta\mu u+\beta^2+16)}{\beta^2\mu^2u^2-8\beta\mu u+\beta^2+16}\right).
\label{taubnutsol2P1}
\end{equation}
\begin{equation}
P_2^2(u)=\left(\frac{\beta^2\mu^2u^2-8\beta\mu u+\beta^2+16}{\mu\beta^2}\right),
\label{taubnutsol2P2}
\end{equation}
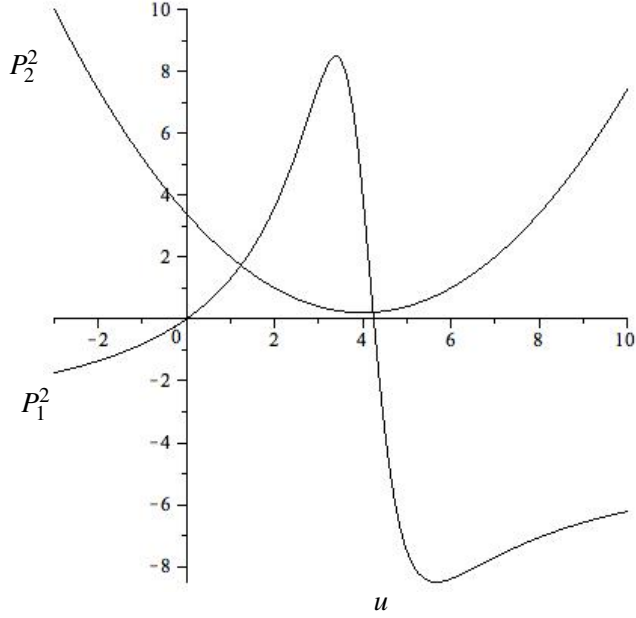
\begin{figure}
\tikzstyle{background grid}=[draw, black!50,step=.5cm]
\begin{tikzpicture}
\node (img) [inner sep=0pt,above right]
{\includegraphics[width=8cm]{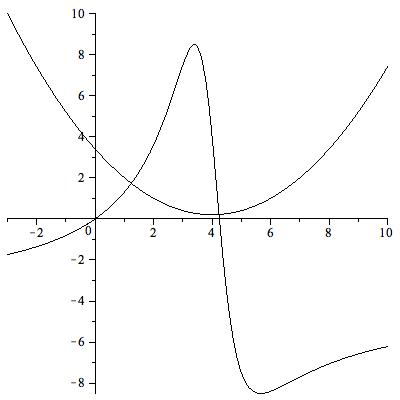}};
\draw (-2:4.5cm) node {$u$};
\draw (92:2.5cm) node {$P_1^2$};
\draw (92:7cm) node {$P_2^2$};
\end{tikzpicture}
\caption[dummy5]{Analytic Taub-NUT solution in terms of $P_1$ and
  $P_2$ for $\mu=\beta=1$.}
\label{figTaubNUT2}
\end{figure}
Note the similarity with the derivation in Section \ref{DLHvac}. We
could now eliminate the integration constants in favour of $a_0$ and
$b_0$, or rather their equivalent series initial conditions.
The functions $P_1(u)$ and $P_2(u)$ are plotted for $\mu=\beta=1$ in
Fig.~\ref{figTaubNUT2}.

\subsubsection{Comparison of the vacuum solution in different set-ups}

It is clear from comparing (\ref{lineelmttaubnut2}) with
(\ref{lineelmttaubnut3}) that 
$g$ and $P^2_1$ are in correspondence
\begin{equation}
\frac{1}{2}P_1^2(u)\sim g(u),
\label{P1g}
\end{equation}
and that the degenerate radii are straightforwardly related as
\begin{equation}
\frac{1}{2}P_2(u)\sim e^{\zeta(u)},
\label{P2ezeta}
\end{equation}
where the $\sim$ is taken to denote equivalence up to the above shift $u_0$.


One sees that applying the above identifications (\ref{P2ezeta}) and
(\ref{P1g}) to the Taub-NUT scaling family (\ref{TNUTscaling}), one
obtains a family of solutions even in the conventional setup:
\begin{equation}
\bar{g}(u)=\frac{1}{\sqrt{\beta}}g(\beta u),\quad
\bar{\zeta}(u)= \zeta(\beta u) - \frac{1}{2}\ln \beta.
\label{RSscaling}
\end{equation}
This can also be seen from changing the time parameter in the metric
itself as before. However, this transform looks less natural than the one for DLH, 
adding to the suspicion that
Taub-NUT time is not a physically sensible coordinate.

One may continue to match the vacuum solution in the three different
set-ups analytically. The above matching of the radii in the two
versions of Taub-NUT together with the relative time shift $u_0$ allow
one to identify $A$ and $B$ in terms of $\mu$ and $\beta$.
This now completes the identification of the two Taub-NUT
versions. These two equivalent models can now in turn be related to
the initial conditions $a_0$ and $b_0$ in the (original and periodic)
DLH model using the time reparameterisation (\ref{timerep2}).

\begin{figure}
\tikzstyle{background grid}=[draw, black!50,step=.5cm]
\begin{tikzpicture}
\node (img) [inner sep=0pt,above right]
{\includegraphics[width=8cm]{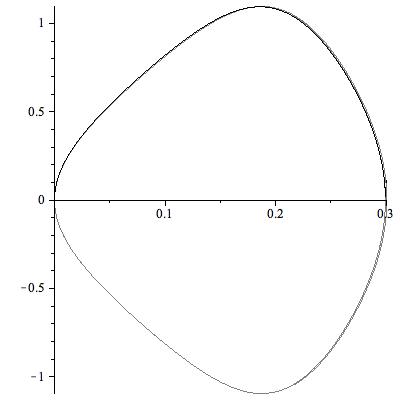}};
\draw (35:5.5cm) node {$u$ and $u(t)$};
\draw (100:4.5cm) node {$\sqrt{g(u)}$,};
\draw (100:3.9cm) node {$P_1(u)$,};
\draw (100:3.4cm) node {$R_1(t)$};
\end{tikzpicture}
\caption[dummy5]{Explicit matching between conventional Taub-NUT
  ($\sqrt{g(u)}$, dashed black), our reparameterised version of
  Taub-NUT ($P_1(u)$, dotted black) and the periodic vacuum DLH model
  mapped over using the diffeomorphism ($R_1(t)$ versus $u(t)$ as
  given by (\ref{timerep2}), continuous gray). They are found to lie on
  top of each other, with the Taub-NUT models covering the upper half
  plane, corresponding to the region where $g(u)\sim P_1^2(u)>0$,
  whereas the DLH vacuum model winds around a closed curve in $u$
  space as it cycles through its periodic behaviour. The parameters
  are matched such that the periodic DLH model has the same initial
  conditions as considered earlier. The other radii $e^\zeta$, $P_2$
  and $R_2$ can be shown to match likewise.}
\label{figMatch}
\end{figure}
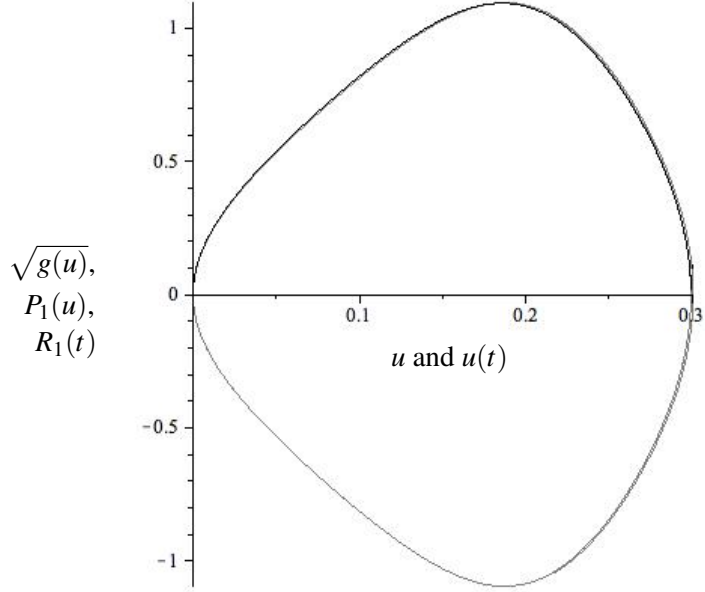

In fact, all three models can be shown to coincide, at least for the
portion where the radius in Taub-NUT corresponds to some physical
distance ($g(u)\sim P_1^2(u)>0$), i.e. where taking the square root
gives something real.  Fig. \ref{figMatch} displays $\sqrt{g}\sim P_1$
from their analytical forms (\ref{taubnutsol}) and
(\ref{taubnutsol2P1}) in the range where $g(u)\sim P_1^2>0$ (
$\sqrt{g}$ dashed, $P_1$ dotted).  They are found to be in agreement
with our matching of these two solutions analytically above.
Furthermore, they also coincide with a plot of $R_1(t)=P_1(u)$ as a
function of $u(t)$ as defined by (\ref{timerep2}), evaluated for the
periodic DLH vacuum solution considered above. However, whereas
$\sqrt{g}$ and $P_1$ only live in the upper half plane, the mapped DLH
solution winds around a mirror-symmetric closed parametric curve in
$P_1$-$u(t)$-space as it cycles through its periodic oscillations,
with one half-period of the elliptic solution corresponding to the
positive portion of $g$. The other radii $e^\zeta$, $P_2$ and $R_2$
match likewise.

\begin{figure}
\tikzstyle{background grid}=[draw, black!50,step=.5cm]
\begin{tikzpicture}
\node (img) [inner sep=0pt,above right]
{\includegraphics[width=8cm]{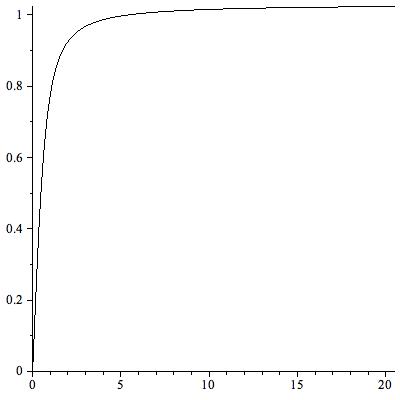}};
\draw (-2:4.5cm) node {$t(u)\sim i u$};
\draw (100:5.5cm) node {$-i{P_1}\sim-i{R_1}$};
\end{tikzpicture}
\caption[dummy5]{First NUT region in Taub-NUT (corresponding to $u<0$) mapped over to DLH space}
\label{figleftNUT}
\end{figure}

This now raises the question: to what do the parts of the analytic
Taub-NUT solutions with $g<0$ correspond when mapped over to the DLH
setup using the inverse mapping (\ref{timerep3})? Clearly, in this
regime, the physical radius $P_1=R_1$ must become imaginary. The
prescription for finding $t(u)$ then indicates that time $t$ must
likewise become imaginary, as follows.
\begin{figure}
\tikzstyle{background grid}=[draw, black!50,step=.5cm]
\begin{tikzpicture}
\node (img) [inner sep=0pt,above right]
{\includegraphics[width=8cm]{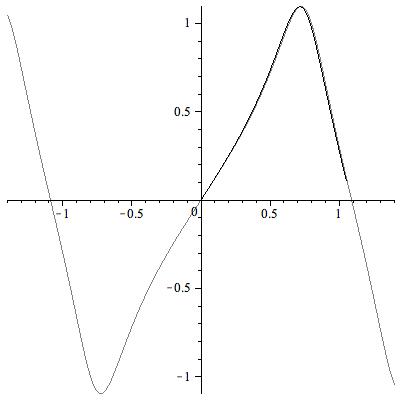}};
\draw (-2:4.5cm) node {$t(u)$};
\draw (100:5.5cm) node {${P_1}\sim{R_1}$};
\end{tikzpicture}
\caption[dummy5]{The Taub region in Taub-NUT (corresponding to $g>0$) mapped over to DLH space (black) falls on top
of the the first half-period of the periodic vacuum DLH solution (grey).}
\label{figtaub}
\end{figure}
Using the relation (\ref{dudt}) together with the known analytic
solutions (\ref{taubnutsol}) and (\ref{taubnutsol2P1}) we find
\begin{equation}
\frac{dt}{du}=\frac{2}{P_1}=\frac{2}{\sqrt{g}}=2\sqrt{\frac{\beta^2\mu^2u^2-8\beta\mu u+\beta^2+16}{\beta u(-4\beta\mu u+\beta^2+16)}} =2\sqrt{\frac{{B(4B^2u^2+1)}}{{Au+1-4B^2u^2}}}.
\label{dtdu2}
\end{equation}
This is well-defined when $g>0$ and just recovers the positive
half-period of the DLH vacuum model, as is obtained numerically (see
Fig. \ref{figtaub}). However, when $g<0$ we can perform a Wick
rotation to imaginary time $\tau=iu$. In the limit of large $u$ where
the Taub-NUT solution approaches a negative constant, we find that $t$
and $u$ are linearly related, but with a relative factor of $i$, such
that $t\sim \tau \sim iu$. So the NUT regions of Taub-NUT correspond
to both imaginary space and time coordinates in  the DLH
model. Fig. \ref{figleftNUT} displays the $u<0$ NUT region mapped
over to DLH by integrating (\ref{dtdu2}) numerically. Here, we have
chosen to plot the result in the first quadrant, as there is a freedom
of choosing factors of $\pm i$ on both axes. At late times, the linear
relationship between $t$ and $u$ means that this mapped NUT region
also settles down to a constant like in the original Taub-NUT
model. So, in particular, mapping the Taub-NUT model does not recover
the periodic DLH solution, and instead corresponds to imaginary space
and time. Performing the integration in (\ref{dtdu2}) analytically
yields a solution in terms of elliptic integrals of precisely the same
form as (\ref{ellintsol2}), as one would expect.

One marked difference between the DLH vacuum solution and this new
version of the Ryan and Shepley form of the Taub-NUT solution is now
that the latter is not periodic, which is the opposite of what one
would commonly expect: normally Lorentzian models are not periodic,
but upon euclideanising become periodic (cf.\ the periodicity in
imaginary time of Green's and partition functions, in particular in
relation to black hole thermodynamics). However,  a similar
phenomenon has recently also been observed in Bianchi V
\cite{Valent2009BianchiVEllipticLorentzian}.

Note that when one allows both the positive and the negative branches 
of $\sqrt{g}$, one recovers precisely the trajectory in $u$-$P_1$-space
that the mapped DLH model traces out (Fig. \ref{figMatch}). However, in the conventional setup
Taub-NUT only selects one branch, and then turns to imaginary time and space on 
either side (the NUT regions, see Fig. \ref{figleftNUT}). This suggests that 
even the Taub-NUT model could cycle indefinitely (like the vacuum DLH model) if
it was allowed to change from one branch to the other. In fact, this matches smoothly:
we have chosen $P_1^2$ to have a zero at the origin, so 
it is linear  there from (\ref{taubnutsol2P1}) . Thus 
$P_1$ goes like $\sqrt{u}$ at the origin, as also observed in Sections \ref{MapDLH} 
(Fig. \ref{figTNUTpancP1}) and \ref{TNUTSF}.  Hence, there is actually a rather natural, smooth 
transition between the two branches, exactly like in Fig. \ref{figTNUTpancP1}. This also suggests that
 the smooth, single-valued parabolic behaviour of DLH is more natural and more fundamental, as opposed to
 the Taub-NUT case, where the parabola essentially gets turned sideways such that the description in terms of
square roots  results in multivaluedness by having the two different branches.


\subsection{Taub-NUT with a scalar field\label{TNUTSF}}

\begin{figure}
\tikzstyle{background grid}=[draw, black!50,step=.5cm]
\begin{tikzpicture}
\node (img) [inner sep=0pt,above right]
{\includegraphics[width=8cm]{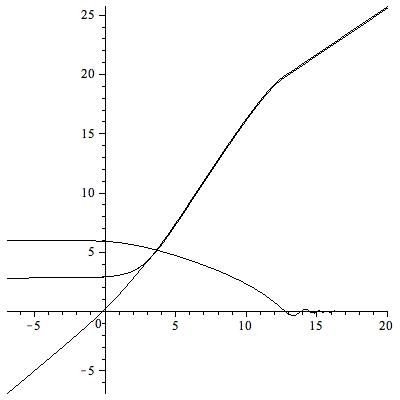}};
\draw (-2:4.5cm) node {$\ln u$};
\draw (100:0.5cm) node {$\ln P_1^2$};
\draw (97:2.7cm) node {$\ln P_2^2$};
\draw (94:3.5cm) node {$F$};
\end{tikzpicture}
\caption[dummy5]{Taub-NUT inflation: inflation and isotropisation in
  analogy to the DLH scenario for the model with parameters
$a_0=1$, $b_0=17$, $\kappa=1$, $\Lambda=0$, $f_0=6$, $m=\frac{1}{8}$.}
\label{figTNUTSF}
\end{figure}
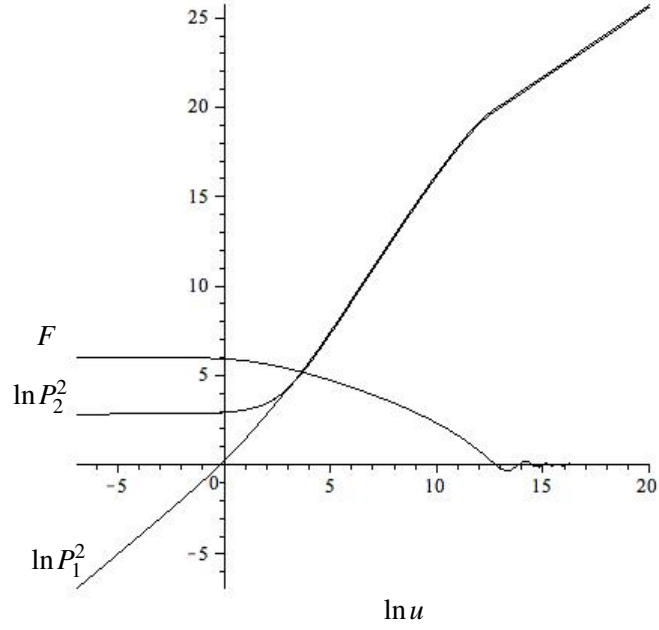

Now that we have clarified the relationship between the DLH and
reparameterised Taub-NUT set-ups in the vacuum case, we can use the
transformation (\ref{timerep2}) to map our cosmological DLH solution
with a scalar field, presented in in
\cite{DechantHobsonLasenby2009BianchiIXScalarField}, directly over to
the reparameterised Taub-NUT model. Alternatively we could find an
analogous series expansion to (\ref{TNUT1})-(\ref{TNUTFr}) directly.

In either case, to lowest order, the series solution has the form
\begin{equation}
P_1(u)=\sqrt{u},\,\,\, P_2(u)=\const, \,\,\,F(u)=\const,
\label{TNUTseries}
\end{equation}
and thus analogous boundary conditions to those used in the DLH case
may be used, from which a numerical integration can be performed
straightforwardly.

\begin{figure}
\tikzstyle{background grid}=[draw, black!50,step=.5cm]
\begin{tikzpicture}
\node (img) [inner sep=0pt,above right]
{\includegraphics[width=8cm]{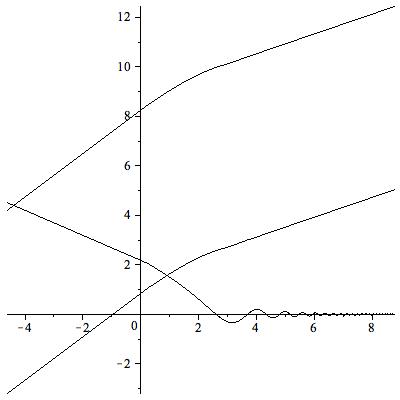}};
\draw (-2:4.5cm) node {$\ln u$};
\draw (100:0.5cm) node {$\ln P_1$};
\draw (97:3.5cm) node {$F$};
\draw (94:5cm) node {$\ln P_2$};
\end{tikzpicture}
\caption[dummy5]{Taub-NUT inflation: inflation, but not complete
  isotropisation for the model with parameters $a_0=1$, $b_0=2$,
  $\kappa=1$, $\Lambda=0$, $f_0=6$, $m=1$.}
\label{figTNUTSF2}
\end{figure}
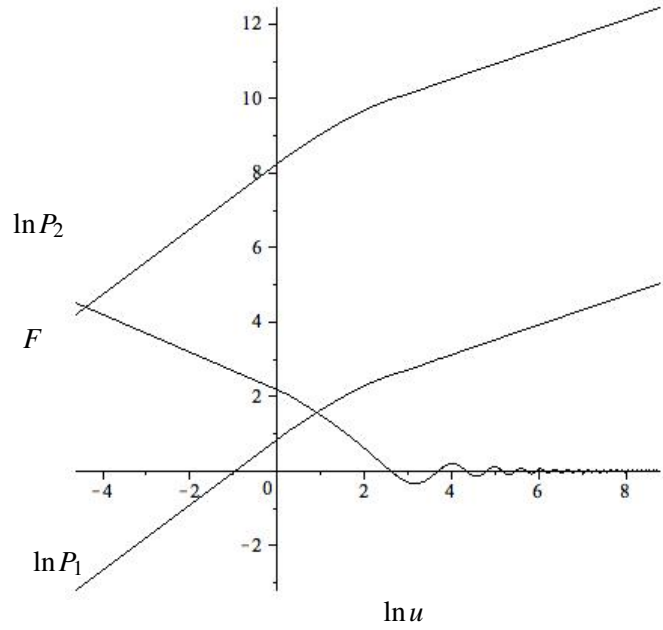

Gradually introducing a light or diffuse scalar field has a similar
effect to the analogous scenario in DLH (Section \ref{DLH_SF}), where
it slightly perturbs the vacuum case.  Fig. \ref{figTNUTSF} shows a
plot of an interesting choice of initial parameters with a sizable
scalar field ($a_0=1.0$, $b_0=17$, $\kappa=1$, $\Lambda=0$, $f_0=6$, 
$m=\frac{1}{8}$), that exhibits inflation
and isotropisation thereby looking rather like the original DLH
solution. Fig. \ref{figTNUTSF2} shows a plot of another interesting
model with the set of initial parameters $a_0=1.0$, $b_0=2.0$,
$\kappa=1$, $\Lambda=0$, $f_0=6$, $m=1$), that exhibits
inflation, but does not completely isotropise. However, the different
Hubble factors tend to a common value so in the flat-space late-time
limit the difference in scale factors would be unobservable as argued
in \cite{DechantHobsonLasenby2009BianchiIXScalarField}.

The slope of $P^2_1$ in Fig. \ref{figTNUTSF} at the pancake is unity,
so that slope $P_1$ here is 1/2, in agreement with the DLH series
solution mapped to Taub-NUT at this point (see equations
(\ref{TNUTseries}) and (\ref{pancakeP1})).  The solution in
Fig. \ref{figTNUTSF2} does not appear to be of the pancaking type, so
we do not consider its early slope here. However, the late-time slopes
in $P_1$ are $0.4$ in both models (i.e. $0.8$ in $P_2^2$). Note, that
this does not agree with the behaviour $a(t)\propto t^{\frac{2}{3}}$ that one
would expect once the scalar field has decayed and behaves like
non-relativistic dust in an Einstein-de-Sitter phase (at which point,
presumably, reheating would occur).  Note, however, that the DLH model
in Fig. \ref{figDLH} does have the required late time slope of
$\frac{2}{3}$, so that the DLH coordinate time matches onto physical
time at late times.  So we conclude that the Taub-NUT time coordinate
is not a physical time coordinate at late times, unlike the DLH time.
This is consistent with the coordinate transformation
(\ref{timerep2}).

We can compare the slopes in the DLH model and the corresponding
Taub-NUT model for late-time power-law behaviour $R(t)=\alpha t^\beta$
as follows. The corresponding slope in Taub-NUT is
\begin{equation}
\frac{dP}{du}=\frac{dR}{dt}\frac{dt}{du}=\frac{dR}{dt}\frac{2}{R}=\frac{2\beta}{t}.
\label{TNUTslope}
\end{equation}
When we can assume that the integral in (\ref{timerep2}) is in fact
dominated by the contribution from the power-law behaviour we can
integrate to get
\begin{equation}
u=\frac{\alpha}{2(\beta+1)}t^{\beta+1} \Rightarrow t=\left(\frac{2(\beta+1)}{\alpha}u\right)^{\frac{1}{\beta+1}}.
\label{TNUTslope2}
\end{equation}
Substituting (\ref{TNUTslope2}) into (\ref{TNUTslope}) then yields
\begin{equation}
\frac{dP}{du}=2\beta\left(\frac{2(\beta+1)}{\alpha}u\right)^{-\frac{1}{\beta+1}}.
\label{TNUTslope3}
\end{equation}
Integrating we obtain
\begin{equation}
P(u)=\alpha^{\frac{1}{\beta+1}}\left(2(\beta+1)\right)^{\frac{\beta}{\beta+1}}u^{\frac{\beta}{\beta+1}}+c.
\label{TNUTslope4}
\end{equation}
For an Einstein-de-Sitter phase with $\beta=\frac{2}{3}$ this therefore gives $P(u)\sim u^{2/5}$, which is indeed what is observed in Figs. \ref{figTNUTSF} and \ref{figTNUTSF2}.

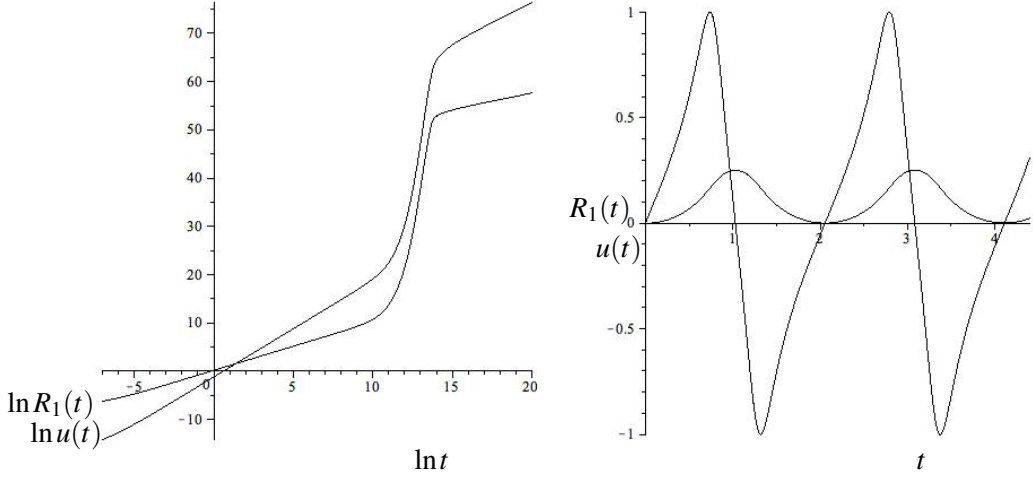
\begin{figure}
\begin{tabular}{@{}c@{ }}
	\tikzstyle{background grid}=[draw, black!50,step=.5cm]
	\begin{tikzpicture}
	\node (img) [inner sep=0pt,above right]
	{\includegraphics[width=6cm]{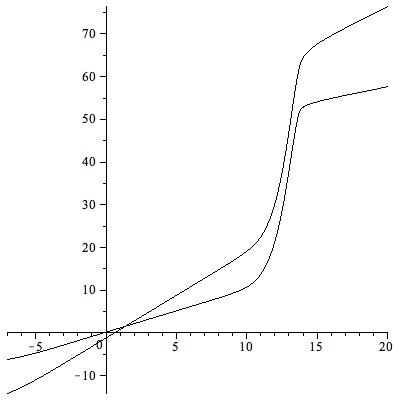}};
	\draw (-2:4.5cm) node {$\ln t$};
	\draw (155:0.4cm) node {$\ln u(t)$};
	\draw (135:0.8cm) node {$\ln R_1(t)$};
	\end{tikzpicture}
    \tikzstyle{background grid}=[draw, black!50,step=.5cm]
	\begin{tikzpicture}
	\node (img) [inner sep=0pt,above right]
	{\includegraphics[width=6cm]{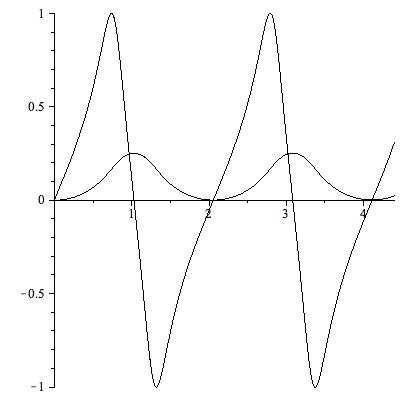}};
	\draw (-2:4.5cm) node {$t$};
	\draw (80:2.7cm) node {$u(t)$};
	\draw (86:3.2cm) node {$R_1(t)$};
	\end{tikzpicture}
  \end{tabular}
\caption[dummy5]{$u(t)$ for the original DLH model with a scalar field
  (left) and the periodic vacuum DLH model (right).}
\label{figslopes}
\end{figure}

Fig. \ref{figslopes} shows $u(t)$ according to (\ref{timerep2}) in
both the original DLH model with a scalar field and the new periodic
vacuum solution. Note that the early time slope in $u(t)$ is $2$, in
agreement with formula (\ref{TNUTslope2}) in the DLH model.  The slope
at late times in the DLH model is $1.67\sim 1+\frac{2}{3}$, also in
agreement with (\ref{TNUTslope2}). Note that the integral
(\ref{timerep2}) is indeed dominated by the contribution after
inflation. As already mentioned in Section \ref{transformation}, in
the periodic model $u(t)$ can be seen to go back on itself when $R_1$
changes sign.  This means that $u$ does not measure the progression of
time in the same way as $t$, but instead retraces a parameter range
that has been traversed previously. Considering that DLH time $t$
matches onto physical time, this casts further strong doubts on
Taub-NUT `time' $u$ being a sensible measure of time.

We conclude that the desirable features of inflation and
isotropisation can survive the mapping to Taub-NUT. However, at early
times -- near the pancaking events -- the DLH coordinatisation seems
much more natural (c.f. Figs \ref{figTNUTpancP1}). At late times, the
physical time coordinate can be inferred from the expansion history of
the universe, and Taub-NUT again fails to match onto something
physical.  Furthermore the scaling solution is most natural in DLH,
and very contrived in Taub-NUT coordinates.  This also ties in with
the fact that the closed Bianchi model is an immediate generalisation
of the physical closed FRW model, which also admits such a scaling
family.  The multivaluedness of the Taub-NUT time coordinate when
mapped from a physically meaningful model is a problem, as is allowing
it to go imaginary in order to reconcile the different solutions.
Thus Taub-NUT time $u$ fails to be a physically sensible measure of
time on several accounts, and we therefore contend that the DLH
coordinates are the more physical coordinate system.

\section{Conclusions\label{conclusion}}
We have compared and contrasted two biaxial Bianchi IX spacetimes --
Taub-NUT and the DLH model. They exhibit great similarities such as
dimensional reduction at a non-singular pancaking event, inspiralling
geodesics and the same eigenvalue structure of the curvature and Weyl
tensors. However, there are profound differences in the global
structure and physical interpretation.  We have shown that the two
metrics can be mapped into each other, but that the coordinate
transformation is multivalued.  This property is responsible for
introducing artifacts into the coordinatisation and thus accounts for
the differences in global behaviour. We believe that our
parameterisation in terms of physical scale factors in complete
analogy with, for instance, FRW-models is more natural, as opposed to
working with the squares thereof and allowing those to become
negative.  In the light of this, we have removed the scalar field from
the original DLH model and found an analytic vacuum solution in terms
of elliptic integrals. This solution is periodic and also recovered by
numerical integration, with which there is very good agreement.  There
is also an obvious link between this periodic solution and the
pancaking series solution of the DLH model, in that the boundary
conditions at the pancake are the same. However, after the inclusion
of the scalar field the DLH model isotropised, inflated and yielded a
viable late-time cosmology with a perturbation spectrum consistent
with observations as shown in
\cite{DechantHobsonLasenby2009BianchiIXScalarField}.  Thus in either
case, working with the scale factors yields sensible cosmological
models and, in particular, the spatial sections are closed Bianchi IX
throughout, in contrast with the topology changing transition in
Taub-NUT.  Mapping the periodic DLH vacuum solution using the
coordinate transformation matches onto the analytic Taub-NUT solution
for the Taub portion where $g>0$. However, in the NUT regions, where
$g<0$, we find that Taub-NUT time corresponds to imaginary time in
`DLH space'.  This firstly adds to the doubts that it is a sensible
time coordinate; secondly, it resolves the alleged topology
transition, as DLH now has closed spatial sections throughout as well
as a physical time coordinate. Thirdly, and rather surprisingly, this
suggests that the Lorentzian Taub-NUT is actually Euclidean in these
regions. It is then surprising that the (Lorentzian) DLH model is
periodic, whereas the `Euclidean Taub-NUT' is not, which is contrary
to what usually happens.  We also note that DLH time essentially acts
as conformal time for Taub-NUT.

The DLH coordinatisation thus has a number of advantages over
traditional approaches. Firstly, the behaviour of the solution across
pancaking events seems much more natural and straightforward in both
the DLH vacuum and scalar field case. The well-behaved odd-parity
series solution that we had found previously acquired bad behaviour
when mapped over to Taub-NUT due to the fact that the transformation
is singular at pancaking events.  Secondly, the periodicity of the DLH
vacuum solution appears more fundamental than the Nut-Taub-Nut
structure and in fact it is clear how this periodicity gets broken by
an integral involved in the mapping. Thirdly, it is clear that the
scale factor has physical significance and its square should therefore
not take negative values. The resulting alternative Taub-NUT
parameterisation still has the advantage of admitting a simple scaling
relation over the conventional metric, which only admits a rather
awkward looking scaling family. Moreover, at late times when we can
infer cosmological time from physical measurements, the DLH model
behaves as one would expect physically, whereas Taub-NUT time again
fails to produce physical results.  We conclude that our
coordinatisation provides at least an interesting alternative point of
view that could shed some light on some of the pathologies that
Taub-NUT is believed to have, and might actually be the more natural
coordinatisation.

\begin{acknowledgements}
{We  thank the  referees for their very useful suggestions.
We would also 
like to thank  Sylvain Br\'echet, Nick Dorey, Gary Gibbons, Jorge Santos, 
Stephen Siklos, David Tong and others
for helpful comments and for pointing out relevant references.
PPD is grateful for support through an STFC (formerly PPARC)
studentship. }
\end{acknowledgements}

\appendix

\section{Mapping of curvature invariants\label{app2}}

Given the similarities in the metric Ansatz, i.e. biaxial Bianchi IX on the hypersurfaces, it was obvious that
similarities between the Taub-NUT and DLH geometries should be explored. In fact, there does exist a coordinate transformation
linking the two spacetimes, which we have presented above. In general, however, when one wishes to examine the relationship of two spacetimes it is often the easier strategy  to show that the two spacetimes are actually distinct. 
This is made difficult by the general coordinate invariance of General Relativity, so one must find coordinate invariant ways to distinguish between spacetimes, such as curvature invariants. A convenient
classification is the Petrov classification, which examines the algebraic properties of the Weyl tensor. This is often
easier than other approaches due to the additional self-duality structure and tracelessness of the Weyl tensor (see, for instance, \cite{GA} for a reference). If, however, it turns out that the
Petrov classification is not enough to distinguish between the two spacetimes (here both are of type D) one must find more  curvature invariants, such as topological invariants, or principal curvatures. If after this kind of tests one has still failed to demonstrate that the two spacetimes one wishes to compare are distinct, then the curvature invariants can be used to find the diffeomorphism linking the two spacetimes, if indeed such a diffeomorphism exists.

The Riemann and Weyl tensors $R_{\mu\nu\sigma\tau}$ and $C_{\mu\nu\sigma\tau}$ are multilinear operators of fourth rank acting on tangent vectors. However, they can also be considered as linear operators acting on bivectors, and as such they  have a characteristic polynomial, whose coefficients and roots (eigenvalues) are polynomial scalar invariants. That is, consider the eigenbivector equations
\begin{equation}
R_{\mu\nu\sigma\tau}S^{\sigma\tau}=\lambda S_{\mu\nu}, \,\,\,\, C_{\mu\nu\sigma\tau}T^{\sigma\tau}=\gamma T_{\mu\nu},
\label{BVeq}
\end{equation}
for some bivectors $S^{\mu\nu}$ and  $T^{\mu\nu}$ and eigenvalues $\lambda$ and $\gamma$ respectively. The degeneracy structure of the eigenvalues
provides a coordinate-invariant way of distinguishing spacetimes. Consideration of the eigenbivector equation of the Weyl tensor
leads to the Petrov classification, which is in general easier to examine due to the additional self-duality structure of the Weyl tensor.
 The eigenvalues of the Riemann  tensor are also known as the `principal curvatures'.

\subsection{Petrov type}
There are six bivectors in $3+1$ dimensions, so that, in general, a fourth rank tensor acting on them has six real eigenvalues. Alternatively,  the self-duality of the Weyl tensor yields a natural complex structure, so that the six real eigenvalues are equivalent to three complex eigenvalues. It turns out that both DLH and Taub-NUT have one degenerate pair of complex eigenvalues, which then also fixes the remaining one due to the tracelessness of the Weyl tensor. This corresponds to Petrov type D, which is what Taub-NUT was previously known to be. In terms of real eigenvalues, there are two degenerate pairs and two singlets. The complex structure also results in
the pairs being complex conjugates of each other, and likewise for the singlets. The real eigenvalues are found to be of the form
\begin{equation}
\gamma^D_1=\gamma^D_3=C_1+\sqrt{C_2}, \,\,
\gamma^D_2=\gamma^D_4=C_1-\sqrt{C_2},\,\,
\gamma^D_5=C_3+\sqrt{C_4},\,\,
\gamma^D_6=C_3-\sqrt{C_4},
\end{equation}
for the DLH model, with some expressions $C_i$ in terms of the $R_i$s and their derivatives. There is one constraint among them since the Weyl tensor is traceless, but we have suppressed the exact results in order to avoid unnecessary clutter. Analogous results hold for the Taub-NUT model for some constants $D_i$, from whence the eigenvalue structure (Petrov D) and tracelessness can be seen, and the two sets of results map into each other as suggested.  

Alternatively, one can consider the Weyl tensor as a function $W$ acting on bivectors $B$, such that the eigenvalue equation is $W(B)=\gamma B$. The self-duality translates into $W(IB)=IW(B)$ for the complex structure denoted by $I$.  In terms of complex eigenvalues, the eigenvalue equation becomes $W(B_i)=(\alpha_i+\beta_i I)B_i$. Solving this in the
DLH and Taub-NUT setups yields the eigenvalues
\begin{equation}
\alpha^D_3=-2\alpha^D_1=-2\alpha^D_2,\,\,
\alpha^T_3=-2\alpha^T_1=-2\alpha^T_2,\,\,
\beta^D_3=-2\beta^D_1=-2\beta^D_2,\,\,
\beta^T_3=-2\beta^T_1=-2\beta^T_2,
\end{equation}
which also enjoy the appropriate degeneracy structure, mapping properties and tracelessness.

\subsection{Principal curvatures}
The Petrov classification might not be sufficient to distinguish between two spacetimes, in which case one can consider
the eigenvalues of the full Riemann tensor, which are also called `principal curvatures'.
These are more complicated, as the self-duality (and therefore natural
complex structure) and tracelessness are lost. However, it again turns out that both models have two degenerate pairs
of real eigenvalues, and two singlets, mirroring the structure of the Weyl tensor. Computation of the eigenvalues gives the following form
\begin{equation}
\lambda^D_1=\lambda^D_3=E_1+\sqrt{E_2},\,\,
\lambda^D_2=\lambda^D_4=E_1-\sqrt{E_2},\,\,
\lambda^D_5= E_3+\sqrt{E_4},\,\,
\lambda^D_6= E_3-\sqrt{E_4},
\end{equation}
for the DLH model, again with analogous results
for the Taub-NUT case. It can be seen that these share the same degeneracy structure of the eigenvalues, which is similar to the one of the Weyl tensor. The two models can also be explicitly mapped into each other, as suggested.

\section{Mapping of geodesics\label{app3}}

In order to complete the mapping between DLH and Taub-NUT, we now turn to 
 showing the equivalence of  the geodesic equations in both models. 
The geodesic equations are most easily obtained using the Lagrangian
formalism, in which
\begin{equation}\label{GXL}
L = {  4}g_{\mu\nu}\dot{x}^\mu \dot{x}^\nu
\end{equation}
is varied with respect to the coordinates $[x^\mu]=(t, x, y, z)$; here
a dot denotes a derivative with respect to some affine parameter
$\lambda$ and the factor of $4$ is included for later
convenience. An explicit realisation of the Maurer-Cartan forms is, for instance, 
\begin{eqnarray}
\omega^{1} & \equiv & dx+\sin y \,dz,\nonumber \\
\omega^{2} & \equiv & \cos x\,dy-\sin x\cos y\,dz,\nonumber \\
\omega^{3} & \equiv & \sin x\,dy+\cos x\cos y\,dz.
\label{inv1frms}
\end{eqnarray}

Inserting the DLH metric (\ref{lineelmtDLH2}) into
$L$ yields
 \begin{equation}
\label{GXLex}
L^D=4\dot{t}^2-R_1^2\dot{x}^2-2R_1^2\sin y \dot{x}\dot{z}-R_2^2\dot{y}^2
-\left(R_1^2\sin^2y+R_2^2\cos^2y
\right)\dot{z}^2.
\end{equation}
Since $L$ is independent of the $x$- and $z$-coordinates, the
corresponding Euler--Lagrange equations yield two conserved quantities
$K_x$ and $K_z$ according to the relations
\begin{eqnarray}
-2R_1^2\left(\dot{x}+\sin y \dot{z}\right) & \equiv & K_x, \label{Kx} \\
-2\left[R_1^2 \sin y\left(\dot{x}+\sin y \dot{z}\right)+R_2^2 \cos^2y
\dot{z}\right] & \equiv & K_z, \label{Kz}
\end{eqnarray}
which may be solved for $\dot{x}$ and $\dot{z}$ to yield
\begin{eqnarray}
    \dot{x} & = & \frac{K_zR_1^2 \sin y -K_x\left(R_1^2\sin^2y+R_2^2\cos^2y
    \right)}{2R_1^2 R_2^2 \cos^2 y}, \label{xdot}\\
\dot{z}&=&\frac{K_x\sin y  - K_z}{2R_2^2 \cos^2 y}. \label{zdot}
\end{eqnarray}

Substituting these expressions back into the Lagrangian, we get the modified form
 \begin{equation}
\label{GXLexDM}
L^D=\frac{1}{4}\,{\frac {16\, \dot{t} ^{2} {R}_1 ^{2}   {R}_2 ^{2} \cos^{2} y   -4\, \dot{y} ^{2}{R}_1 ^{2}{R}_2 ^{4}   \cos^{2} y   -\left({K_x}^{2}+{K_z}^{2}\right) {R}_1 ^{2}+2\, {R}_1 ^{2}\sin y K_xK_z+{K_x}^{2} \left({R}_1 ^{2}  - {R}_2 ^{2}\right)\cos^{2} y  } {{R}_1 ^{2}  {R}_2 ^{2} \cos^{2} y     }}.
\end{equation}
The Euler--Lagrange equation for $y$ then reads
\begin{equation}\label{ELy}
    4R_2^3 \cos^3y\left(R_2\ddot{y}+2\frac{\partial R_2}{\partial t}\dot{t}\dot{y}\right)=
    \left(K_x^2+K_z^2\right)\sin y -2K_xK_z+K_xK_z \cos^2 y,
\end{equation}
as stated and further analysed in \cite{DechantHobsonLasenby2009BianchiIXScalarField}.

For Taub-NUT we keep the same variables for now, except $t$ is replaced by $u$. (In general, the 1-forms would be of the same form, but the coordinates and the affine parameter could be different.) We will see later that this choice is indeed consistent. 
\begin{equation}\label{GXLT}
L = {  4}g_{\mu\nu}\dot{x}^\mu \dot{x}^\nu
\end{equation}
is varied with respect to the coordinates $[x^\mu]=(u, x, y, z)$. Inserting the Taub-NUT metric in the form (\ref{lineelmttaubnut3}) into
$L$ yields
 \begin{equation}
\label{GXLexT}
L^T=8\dot{u}\left(\dot{x}+\sin y \dot{z}\right)-P_1^2\dot{x}^2-2P_1^2\sin y \dot{x}\dot{z}-P_2^2\dot{y}^2
-\left(P_1^2\sin^2y+P_2^2\cos^2y\right)\dot{z}^2.
\end{equation}

Since this Langrangian is likewise independent of the $x$- and $z$-coordinates, we still have the
 two conserved quantities $K_x$ and $K_z$. However, they pick up an extra term in $\dot{u}$
\begin{eqnarray}
8\dot{u}-2P_1^2\left(\dot{x}+\sin y \dot{z}\right) & \equiv & K_x, \label{KxTx} \\
8\dot{u}\sin y-2\left[P_1^2 \sin y\left(\dot{x}+\sin y \dot{z}\right)+P_2^2 \cos^2y
\dot{z}\right] & \equiv & K_z \equiv K_x \sin y-2P_2^2 \cos^2y
\dot{z}, \label{KzT}
\end{eqnarray}
which may be solved for $\dot{x}$ and $\dot{z}$ to yield
\begin{eqnarray}
    \dot{x} & = & \frac{4\dot{u}}{P_1^2}+\frac{K_zP_1^2 \sin y -K_x\left(P_1^2\sin^2y+P_2^2\cos^2y
    \right)}{2P_1^2 P_2^2 \cos^2 y}, \label{xdotT}\\
\dot{z}&=&\frac{K_x\sin y  - K_z}{2P_2^2 \cos^2 y}. \label{zdotT}
\end{eqnarray}
Substituting these expressions back into the Lagrangian, we get the modified form
 \begin{equation}
\label{GXLexTM}
L^T=\frac{1}{4}\,{\frac {64\,  \dot{u}    ^{2} {P}_2 ^{2}  \cos^{2} y  -4\, \dot{y}   ^{2}{P}_1 ^{2}{P}_2 ^{4}    \cos^{2} y   -\left({K_x}^{2}+{K_z}^{2} \right) {P}_1 ^{2} +2\, {P}_1 ^{2}\sin y K_{x}K_z+{K_x}^{2} \left( P_1 ^{2}-   {P}_2 ^{2}\right) \cos ^{2}  y     }{ {P}_1 ^{2}{P}_2 ^{2}  \cos^{2} y   }},
\end{equation}
where the terms that are linear in $\dot{u}$ have cancelled, and the use of the conserved quantities has in fact introduced a term that is quadratic in $\dot{u}$.

Comparing with (\ref{GXLexDM}), we see that a straightforward identification of coordinates, scale factors and affine parameters is indeed possible, and the differing terms $64\,  \dot{u}    ^{2} {P}_2 ^{2}  \cos^{2} y$ and $16\, \dot{t} ^{2} {R}_1 ^{2}   {R}_2 ^{2} \cos^{2} y $ are consistent with the time rescaling (\ref{dudt}). Therefore the equivalence of the geodesics is already exhibited at the level of the Lagrangians. The Euler-Lagrange equation for $y$ in Taub-NUT is indeed identical to (\ref{ELy}), as the Lagrangians only differ on the time-time piece, which is independent of $y$ and $\dot{y}$.

\section{Derivation of the elliptic integral solution for
the vacuum DLH model\label{app1}}

In order to recover the analytic form of the numerical solution in
Fig. \ref{figDLHvacperiodic2} we note that an appropriate combination
of (\ref{vacspecial1}) and (\ref{vacspecialFr}) allows us to integrate
a Bernoulli-type equation in $R_1(t)^{-2}$ to find $R_1$ in terms of
$R_2$ and an integration constant $\mu$ as
\begin{equation}
{R_1(t)=\pm\frac{R_2(t)\dot{R}_2(t)}{\sqrt{\mu R_2(t)^2-1}}}.
\label{R1itoR2}
\end{equation}
As will become clear momentarily, we will choose the minus sign, in
order to recover the analytic form of the numerical solution
above. Substituting this expression for $R_1$ into equation
(\ref{vacspecial1}) allows us in turn to solve for $R_2$. This gives
another integral solution with positive and negative branches and two
integration constants. Differentiating and choosing the branch that is
decreasing at $t=0$, we lose one integration constant and can fix the
other in terms of $\mu$ and $b_0$ such that $\dot{R}_2(0)$ vanishes in
line with the pancaking series solution and the numerical solution
above.  From (\ref{R1itoR2}), $R_1$ will then be increasing and vanish
at $t=0$ as asserted earlier.  The slope of $R_1$, which is by
definition $a_0$, can then be used to fix $\mu$ and hence the whole
solution in terms of initial conditions
\begin{equation}
\mu=\frac{a_0^2+4}{a_0^2b_0^2}.
\label{mu}
\end{equation}
 With the constants in the problem explicitly fixed, we now return to
 the integration of $R_2$ and $R_1$.  From the denominator in
 (\ref{R1itoR2}) this will involve finding integrals of the type
\begin{equation}
 \sqrt{a_0^2R_2(t)^2+4R_2(t)^2-a_0^2b_0^2}\equiv x.
\label{musub}
\end{equation}
In fact it can be easily seen from the numerical solution that $b_0$
simply scales both axes, so we set it to unity in the following.
\begin{figure}
\tikzstyle{background grid}=[draw, black!50,step=.5cm]
\begin{tikzpicture}
\node (img) [inner sep=0pt,above right]
{\includegraphics[width=8cm]{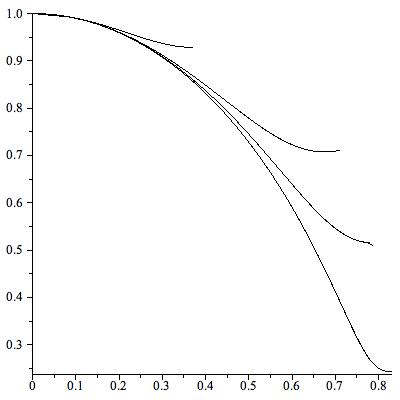}};
\draw (-2:4.5cm) node {$t$};
\draw (100:5cm) node {$R_2$};
\draw (4:8.7cm) node {$a_0=0.5$};
\draw (20:9cm) node {$a_0=1.2$};
\draw (34:9cm) node {$a_0=2$};
\draw (57:8.5cm) node {$a_0=5$};
\end{tikzpicture}
\caption[dummy5]{Comparison of analytic and numerical solutions (which
  coincide) for 4 values of $a_0$: $a_0=0.5,\,\,\, 1.2,\,\,\, 2,\,\,\,
  5$. For each of these values the analytic expression is indeed real
  and coincides with the numerical result to great accuracy.  Note
  that due to the critical factors of $\sqrt{a_0-2}$ in
  (\ref{ellintsol}) there can be relative signs depending on which
  branch one chooses in the function definitions as one crosses
  $a_0=2$. Here, the last two terms have flipped signs for $a_0<2$
  when explicitly evaluated (using Maple) relative to the analytical
  ones in (\ref{ellintsol}). Time $t$ is calculated as a function of
  $R_2$ in the region between the initial maximum and the first
  minimum of $R_2$ due to the multivaluedness and then plotted along
  the horizontal for comparison with the numerical results. }
\label{figDLHvacperanalnum}
\end{figure}
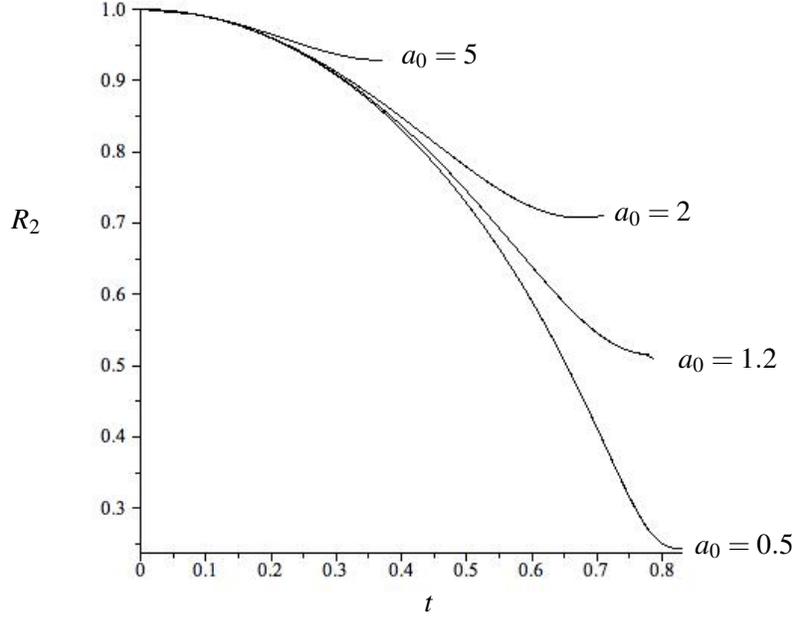
Inverting (\ref{musub}) and substituting in for $R_2$ in terms of $x$
gives a dynamical equation for $x$ that can be integrated to find time
$t$ in terms of $x$ (and thus $R_2$) as
\begin{equation}
t=\sqrt{\frac{i}{2a_0}\frac{x^2+a_0^2}{a_0^2+4}}y
+\sqrt{\frac{a_0}{8i}} 				E\left(y;i\right)
+\sqrt{\frac{ia_0}{32}}(a_0+2i)		F\left(y; i\right) 
-\sqrt{\frac{i}{32a_0}}(a_0^2-4)	\Pi\left(y; \frac{2}{ia_0};i\right),
\label{ellintsol}
\end{equation}
where the elliptic integrals have been reduced in terms of Legendre's
three normal forms
\begin{align}
E(z; k) = \int_{0}^z{\frac{\sqrt{1-k^2s^2}}{\sqrt{1-s^2}}ds},\notag\\ 
F(z; k) = \int_{0}^z{\frac{1}{\sqrt{1-s^2}\sqrt{1-k^2s^2}}ds},\\
\Pi(z;\nu;k) = \int_{0}^z{\frac{1}{(1-\nu s^2)\sqrt{1-s^2}\sqrt{1-k^2s^2}}ds}.\notag
\label{ellint}
\end{align}
and $y\equiv\sqrt{ia_0\frac{(x-2)}{(2x+a_0^2)}}$.

Implicit in the Legendre forms is a choice of branches. 
In (\ref{ellintsol}) these are chosen such that Maple's default choice
indeed recovers a real solution as required. 
The  explicit numerical evaluation of this analytic solution in Maple
is confirmed to be real (to within numerical
precision) and Fig. \ref{figDLHvacperanalnum} shows very good
agreement between the analytical solution and the results from the numerical integration
for four values of $a_0$: $a_0=0.5,\,\,\, 1.2,\,\,\, 2,\,\,\, 5$.
Time $t$ has been calculated as a function of $R_2$ between the first
maximum and minimum of $R_2$ due to its multivaluedness, and then
plotted along the horizontal axis for comparison with the numerical
results.

If one repeats the above analysis for $b_0\ne 1$ one finds that $b_0$
always occurs with the same power as $x$ as might be expected from the
definition. Changes in $b_0$ can therefore be absorbed into
$x$. Moreover, the arguments in the elliptic functions turn out to
have overall power $0$ whereas the coefficients have weight one.  This
explains how the time $t$ gets simply rescaled. Of course, $R_2$ also
gets rescaled, given that $b_0$ is its boundary condition at $t=0$
(which is effectively the same as absorbing it into $x$).

\bibliography{DHLTaubNUT_mph.bbl}

\begin{thebibliography}{34}
\expandafter\ifx\csname natexlab\endcsname\relax\def\natexlab#1{#1}\fi
\expandafter\ifx\csname bibnamefont\endcsname\relax
  \def\bibnamefont#1{#1}\fi
\expandafter\ifx\csname bibfnamefont\endcsname\relax
  \def\bibfnamefont#1{#1}\fi
\expandafter\ifx\csname citenamefont\endcsname\relax
  \def\citenamefont#1{#1}\fi
\expandafter\ifx\csname url\endcsname\relax
  \def\url#1{\texttt{#1}}\fi
\expandafter\ifx\csname urlprefix\endcsname\relax\def\urlprefix{URL }\fi
\providecommand{\bibinfo}[2]{#2}
\providecommand{\eprint}[2][]{\url{#2}}

\bibitem[{\citenamefont{Dechant et~al.}(2009)\citenamefont{Dechant, Lasenby,
  and Hobson}}]{DechantHobsonLasenby2009BianchiIXScalarField}
\bibinfo{author}{\bibfnamefont{P.-P.} \bibnamefont{Dechant}},
  \bibinfo{author}{\bibfnamefont{A.~N.} \bibnamefont{Lasenby}},
  \bibnamefont{and} \bibinfo{author}{\bibfnamefont{M.~P.}
  \bibnamefont{Hobson}}, \bibinfo{journal}{Phys. Rev. D}
  \textbf{\bibinfo{volume}{79}}, \bibinfo{eid}{043524} (\bibinfo{year}{2009}).

\bibitem[{\citenamefont{Bianchi}(1898)}]{bianchi1897}
\bibinfo{author}{\bibfnamefont{L.}~\bibnamefont{Bianchi}},
  \bibinfo{journal}{Mem. della Soc. It. delle Scienze}
  \textbf{\bibinfo{volume}{11}}, \bibinfo{pages}{267} (\bibinfo{year}{1898}).

\bibitem[{\citenamefont{Belinskii et~al.}(1972)\citenamefont{Belinskii,
  Khalatnikov, and Lifshitz}}]{BelinskyKhalatnikovLifshitz1972}
\bibinfo{author}{\bibfnamefont{V.~A.} \bibnamefont{Belinskii}},
  \bibinfo{author}{\bibfnamefont{I.~M.} \bibnamefont{Khalatnikov}},
  \bibnamefont{and} \bibinfo{author}{\bibfnamefont{E.~M.}
  \bibnamefont{Lifshitz}}, \bibinfo{journal}{Zh. Eksp. Teor. Fiz.}
  \textbf{\bibinfo{volume}{62}}, \bibinfo{pages}{1606} (\bibinfo{year}{1972}).

\bibitem[{\citenamefont{Belinskii et~al.}(1971)\citenamefont{Belinskii,
  Khalatnikov, and Lifshitz}}]{BelinskyKhalatnikovLifshitzKL1969}
\bibinfo{author}{\bibfnamefont{V.~A.} \bibnamefont{Belinskii}},
  \bibinfo{author}{\bibfnamefont{I.~M.} \bibnamefont{Khalatnikov}},
  \bibnamefont{and} \bibinfo{author}{\bibfnamefont{E.~M.}
  \bibnamefont{Lifshitz}}, \bibinfo{journal}{Zh. Eksp. Teor. Fiz.}
  \textbf{\bibinfo{volume}{60}}, \bibinfo{pages}{1969} (\bibinfo{year}{1971}).

\bibitem[{\citenamefont{Misner}(1969)}]{Misner1969Mixmaster}
\bibinfo{author}{\bibfnamefont{C.~W.} \bibnamefont{Misner}},
  \bibinfo{journal}{Phys. Rev. Lett.} \textbf{\bibinfo{volume}{22}},
  \bibinfo{pages}{1071} (\bibinfo{year}{1969}).

\bibitem[{\citenamefont{Novikov and
  Zel'dovich}(1973)}]{Novikov1973ProcessesNearSingularities}
\bibinfo{author}{\bibfnamefont{I.~D.} \bibnamefont{Novikov}} \bibnamefont{and}
  \bibinfo{author}{\bibfnamefont{Y.~B.} \bibnamefont{Zel'dovich}},
  \bibinfo{journal}{Annu. Rev. Astro. Astrophys.}
  \textbf{\bibinfo{volume}{11}}, \bibinfo{pages}{387} (\bibinfo{year}{1973}).

\bibitem[{\citenamefont{{Ringstr{\"o}m}}(2001)}]{Ringstrom2001BianchiIXattract%
or}
\bibinfo{author}{\bibfnamefont{H.}~\bibnamefont{{Ringstr{\"o}m}}},
  \bibinfo{journal}{Annales Henri Poincar{\'e}, vol.~2, issue 3, pp.~405-500}
  \textbf{\bibinfo{volume}{2}}, \bibinfo{pages}{405} (\bibinfo{year}{2001}).

\bibitem[{\citenamefont{{Heinzle} and
  {Uggla}}(2009{\natexlab{a}})}]{UgglaHeinzle2009mixmaster}
\bibinfo{author}{\bibfnamefont{J.~M.} \bibnamefont{{Heinzle}}}
  \bibnamefont{and} \bibinfo{author}{\bibfnamefont{C.}~\bibnamefont{{Uggla}}},
  \bibinfo{journal}{Classical and Quantum Gravity}
  \textbf{\bibinfo{volume}{26}}, \bibinfo{pages}{075016}
  (\bibinfo{year}{2009}{\natexlab{a}}), \eprint{0901.0776}.

\bibitem[{\citenamefont{{Heinzle} and
  {Uggla}}(2009{\natexlab{b}})}]{UgglaHeinzle2009newproof}
\bibinfo{author}{\bibfnamefont{J.~M.} \bibnamefont{{Heinzle}}}
  \bibnamefont{and} \bibinfo{author}{\bibfnamefont{C.}~\bibnamefont{{Uggla}}},
  \bibinfo{journal}{Classical and Quantum Gravity}
  \textbf{\bibinfo{volume}{26}}, \bibinfo{pages}{075015}
  (\bibinfo{year}{2009}{\natexlab{b}}), \eprint{0901.0806}.

\bibitem[{\citenamefont{{Henneaux}}(2008)}]{Henneaux2008KacMoodyBKL}
\bibinfo{author}{\bibfnamefont{M.}~\bibnamefont{{Henneaux}}},
  \bibinfo{journal}{ArXiv e-prints}  (\bibinfo{year}{2008}),
  \eprint{0806.4670}.

\bibitem[{\citenamefont{{Henneaux} et~al.}(2008)\citenamefont{{Henneaux},
  {Persson}, and
  {Spindel}}}]{HennauxPersson2008SpacelikeSingularitiesAndHiddenSymmetriesofGr%
avity}
\bibinfo{author}{\bibfnamefont{M.}~\bibnamefont{{Henneaux}}},
  \bibinfo{author}{\bibfnamefont{D.}~\bibnamefont{{Persson}}},
  \bibnamefont{and}
  \bibinfo{author}{\bibfnamefont{P.}~\bibnamefont{{Spindel}}},
  \bibinfo{journal}{Living Reviews in Relativity}
  \textbf{\bibinfo{volume}{11}}, \bibinfo{pages}{1} (\bibinfo{year}{2008}),
  \eprint{0710.1818}.

\bibitem[{\citenamefont{Henneaux et~al.}(2008)\citenamefont{Henneaux, Persson,
  and Wesley}}]{HenneauxPerssonWesley2008CoxeterGroupStructure}
\bibinfo{author}{\bibfnamefont{M.}~\bibnamefont{Henneaux}},
  \bibinfo{author}{\bibfnamefont{D.}~\bibnamefont{Persson}}, \bibnamefont{and}
  \bibinfo{author}{\bibfnamefont{D.~H.} \bibnamefont{Wesley}},
  \bibinfo{journal}{Journal of High Energy Physics}
  \textbf{\bibinfo{volume}{2008}}, \bibinfo{pages}{052} (\bibinfo{year}{2008}).

\bibitem[{\citenamefont{{Cornish} and
  {Levin}}(1999)}]{CornishLevin1997unambiguously}
\bibinfo{author}{\bibfnamefont{N.~J.} \bibnamefont{{Cornish}}}
  \bibnamefont{and} \bibinfo{author}{\bibfnamefont{J.~J.}
  \bibnamefont{{Levin}}}, in \emph{\bibinfo{booktitle}{Recent Developments in
  Theoretical and Experimental General Relativity, Gravitation, and
  Relativistic Field Theories}}, edited by
  \bibinfo{editor}{\bibfnamefont{T.}~\bibnamefont{{Piran}}} \bibnamefont{and}
  \bibinfo{editor}{\bibfnamefont{R.}~\bibnamefont{{Ruffini}}}
  (\bibinfo{year}{1999}), pp. \bibinfo{pages}{616--+}.

\bibitem[{\citenamefont{Cornish and
  Levin}(1997{\natexlab{a}})}]{CornishLevin1997MixmasterChaotic}
\bibinfo{author}{\bibfnamefont{N.~J.} \bibnamefont{Cornish}} \bibnamefont{and}
  \bibinfo{author}{\bibfnamefont{J.~J.} \bibnamefont{Levin}},
  \bibinfo{journal}{Phys. Rev. Lett.} \textbf{\bibinfo{volume}{78}},
  \bibinfo{pages}{998} (\bibinfo{year}{1997}{\natexlab{a}}).

\bibitem[{\citenamefont{Cornish and
  Levin}(1997{\natexlab{b}})}]{CornishLevin1997Fareytale}
\bibinfo{author}{\bibfnamefont{N.~J.} \bibnamefont{Cornish}} \bibnamefont{and}
  \bibinfo{author}{\bibfnamefont{J.~J.} \bibnamefont{Levin}},
  \bibinfo{journal}{Phys. Rev. D} \textbf{\bibinfo{volume}{55}},
  \bibinfo{pages}{7489} (\bibinfo{year}{1997}{\natexlab{b}}).

\bibitem[{\citenamefont{Calogero and Heinzle}(2009)}]{Heinzle2009BianchiLRS}
\bibinfo{author}{\bibfnamefont{S.}~\bibnamefont{Calogero}} \bibnamefont{and}
  \bibinfo{author}{\bibfnamefont{J.~M.} \bibnamefont{Heinzle}},
  \bibinfo{journal}{ArXiv e-prints}  (\bibinfo{year}{2009}),
  \eprint{0911.0667}.

\bibitem[{\citenamefont{{Grishchuk} et~al.}(1976)\citenamefont{{Grishchuk},
  {Doroshkevich}, and {Iudin}}}]{Grishchuk1976Bianchi}
\bibinfo{author}{\bibfnamefont{L.~P.} \bibnamefont{{Grishchuk}}},
  \bibinfo{author}{\bibfnamefont{A.~G.} \bibnamefont{{Doroshkevich}}},
  \bibnamefont{and} \bibinfo{author}{\bibfnamefont{V.~M.}
  \bibnamefont{{Iudin}}}, \bibinfo{journal}{Zhurnal Eksperimental noi i
  Teoreticheskoi Fiziki} \textbf{\bibinfo{volume}{69}}, \bibinfo{pages}{1857}
  (\bibinfo{year}{1976}).

\bibitem[{\citenamefont{Lasenby and
  Doran}(2005)}]{LasenbyDoran2005ClosedUniversesdSandInflation}
\bibinfo{author}{\bibfnamefont{A.}~\bibnamefont{Lasenby}} \bibnamefont{and}
  \bibinfo{author}{\bibfnamefont{C.}~\bibnamefont{Doran}},
  \bibinfo{journal}{Phys. Rev. D} \textbf{\bibinfo{volume}{71}},
  \bibinfo{pages}{063502} (\bibinfo{year}{2005}).

\bibitem[{\citenamefont{Ryan and
  Shepley}(1975)}]{RyanShepley1975HomogeneousRelativisticCosmologies}
\bibinfo{author}{\bibfnamefont{M.~P.} \bibnamefont{Ryan}} \bibnamefont{and}
  \bibinfo{author}{\bibfnamefont{L.~C.} \bibnamefont{Shepley}},
  \emph{\bibinfo{title}{Homogeneous Relativistic Cosmologies}}
  (\bibinfo{publisher}{Princeton University Press, Princeton, NJ},
  \bibinfo{year}{1975}).

\bibitem[{\citenamefont{Konkowski et~al.}(1985)\citenamefont{Konkowski,
  Helliwell, and Shepley}}]{KonkowskiHelliwellShepley1985quasiregularI}
\bibinfo{author}{\bibfnamefont{D.~A.} \bibnamefont{Konkowski}},
  \bibinfo{author}{\bibfnamefont{T.~M.} \bibnamefont{Helliwell}},
  \bibnamefont{and} \bibinfo{author}{\bibfnamefont{L.~C.}
  \bibnamefont{Shepley}}, \bibinfo{journal}{Phys. Rev. D}
  \textbf{\bibinfo{volume}{31}}, \bibinfo{pages}{1178} (\bibinfo{year}{1985}).

\bibitem[{\citenamefont{Konkowski and
  Helliwell}(1985)}]{KonkowskiHelliwell1985quasiregularII}
\bibinfo{author}{\bibfnamefont{D.~A.} \bibnamefont{Konkowski}}
  \bibnamefont{and} \bibinfo{author}{\bibfnamefont{T.~M.}
  \bibnamefont{Helliwell}}, \bibinfo{journal}{Phys. Rev. D}
  \textbf{\bibinfo{volume}{31}}, \bibinfo{pages}{1195} (\bibinfo{year}{1985}).

\bibitem[{\citenamefont{Carter}(1967)}]{Carter2006Thesis}
\bibinfo{author}{\bibfnamefont{B.}~\bibnamefont{Carter}}, Ph.D. thesis,
  \bibinfo{school}{University of Cambridge, UK} (\bibinfo{year}{1967}).

\bibitem[{\citenamefont{{Carter}}(1970)}]{Carter1970Commutation}
\bibinfo{author}{\bibfnamefont{B.}~\bibnamefont{{Carter}}},
  \bibinfo{journal}{Communications in Mathematical Physics}
  \textbf{\bibinfo{volume}{17}}, \bibinfo{pages}{233} (\bibinfo{year}{1970}).

\bibitem[{\citenamefont{{Misner} and
  {Taub}}(1969)}]{MisnerTaub1969SingularityFreeEmpty}
\bibinfo{author}{\bibfnamefont{C.~W.} \bibnamefont{{Misner}}} \bibnamefont{and}
  \bibinfo{author}{\bibfnamefont{A.~H.} \bibnamefont{{Taub}}},
  \bibinfo{journal}{Soviet Journal of Experimental and Theoretical Physics}
  \textbf{\bibinfo{volume}{28}}, \bibinfo{pages}{122} (\bibinfo{year}{1969}).

\bibitem[{\citenamefont{Hawking and
  Ellis}(1973)}]{HawkingEllis1973LargeScaleStructureOfSpaceTime}
\bibinfo{author}{\bibfnamefont{S.}~\bibnamefont{Hawking}} \bibnamefont{and}
  \bibinfo{author}{\bibfnamefont{G.}~\bibnamefont{Ellis}},
  \emph{\bibinfo{title}{The Large Scale Structure of Space-Time}}, Cambridge
  Monographs on Mathematical Physics (\bibinfo{publisher}{Cambridge University
  Press}, \bibinfo{address}{Cambridge, U.K.}, \bibinfo{year}{1973}).

\bibitem[{\citenamefont{{Siklos}}(1978)}]{Siklos1978Whimper}
\bibinfo{author}{\bibfnamefont{S.~T.~C.} \bibnamefont{{Siklos}}},
  \bibinfo{journal}{Communications in Mathematical Physics}
  \textbf{\bibinfo{volume}{58}}, \bibinfo{pages}{255} (\bibinfo{year}{1978}).

\bibitem[{\citenamefont{Kagramanova et~al.}(2010)\citenamefont{Kagramanova,
  Kunz, Hackmann, and Laemmerzahl}}]{Laemmerzahl2010TaubNUTgeodesics}
\bibinfo{author}{\bibfnamefont{V.}~\bibnamefont{Kagramanova}},
  \bibinfo{author}{\bibfnamefont{J.}~\bibnamefont{Kunz}},
  \bibinfo{author}{\bibfnamefont{E.}~\bibnamefont{Hackmann}}, \bibnamefont{and}
  \bibinfo{author}{\bibfnamefont{C.}~\bibnamefont{Laemmerzahl}}
  (\bibinfo{year}{2010}), \eprint{1002.4342}.

\bibitem[{\citenamefont{Ellis and
  Schmidt}(1977)}]{EllisSchmidt1977SingularSpacetimes}
\bibinfo{author}{\bibfnamefont{G.~F.~R.} \bibnamefont{Ellis}} \bibnamefont{and}
  \bibinfo{author}{\bibfnamefont{B.~G.} \bibnamefont{Schmidt}},
  \bibinfo{journal}{Gen. Rel. Grav.} \textbf{\bibinfo{volume}{8}},
  \bibinfo{pages}{915} (\bibinfo{year}{1977}).

\bibitem[{\citenamefont{Krasnikov}(2009)}]{Krasnikov2009QuasiregularSingularit%
iesTakenSeriously}
\bibinfo{author}{\bibfnamefont{S.}~\bibnamefont{Krasnikov}},
  \bibinfo{journal}{ArXiv e-prints}  (\bibinfo{year}{2009}),
  \eprint{0909.4963}.

\bibitem[{\citenamefont{{Konkowski} and
  {Helliwell}}(2005)}]{KonkowskiHelliwell2005Classification}
\bibinfo{author}{\bibfnamefont{D.~A.} \bibnamefont{{Konkowski}}}
  \bibnamefont{and} \bibinfo{author}{\bibfnamefont{T.~M.}
  \bibnamefont{{Helliwell}}}, in \emph{\bibinfo{booktitle}{The Tenth Marcel
  Grossmann Meeting. On recent developments in theoretical and experimental
  general relativity, gravitation and relativistic field theories}}, edited by
  \bibinfo{editor}{\bibfnamefont{M.}~\bibnamefont{{Novello}}},
  \bibinfo{editor}{\bibfnamefont{S.}~\bibnamefont{{Perez Bergliaffa}}},
  \bibnamefont{and} \bibinfo{editor}{\bibfnamefont{R.}~\bibnamefont{{Ruffini}}}
  (\bibinfo{year}{2005}), pp. \bibinfo{pages}{1829--+}.

\bibitem[{\citenamefont{Stephani et~al.}(2003)}]{Stephani2003ExactSolutions}
\bibinfo{author}{\bibfnamefont{H.}~\bibnamefont{Stephani}}
  \bibnamefont{et~al.}, \emph{\bibinfo{title}{Exact Solutions to {Einstein}'s
  Field Equations, 2nd edition}} (\bibinfo{publisher}{Cambridge University
  Press}, \bibinfo{year}{2003}).

\bibitem[{\citenamefont{{Pleba\'nski} and
  {Demia\'nski}}(1976)}]{PlebanskiDemianski1976RotatingChargedandUniformlyAcce%
lerating}
\bibinfo{author}{\bibfnamefont{J.~F.} \bibnamefont{{Pleba\'nski}}}
  \bibnamefont{and}
  \bibinfo{author}{\bibfnamefont{M.}~\bibnamefont{{Demia\'nski}}},
  \bibinfo{journal}{Annals of Physics} \textbf{\bibinfo{volume}{98}},
  \bibinfo{pages}{98} (\bibinfo{year}{1976}).

\bibitem[{\citenamefont{{Valent}}(2009)}]{Valent2009BianchiVEllipticLorentzian}
\bibinfo{author}{\bibfnamefont{G.}~\bibnamefont{{Valent}}},
  \bibinfo{journal}{General Relativity and Gravitation}
  \textbf{\bibinfo{volume}{41}}, \bibinfo{pages}{2433} (\bibinfo{year}{2009}),
  \eprint{1002.1454}.

\bibitem[{\citenamefont{Doran and Lasenby}(2003)}]{GA}
\bibinfo{author}{\bibfnamefont{C.}~\bibnamefont{Doran}} \bibnamefont{and}
  \bibinfo{author}{\bibfnamefont{A.~N.} \bibnamefont{Lasenby}},
  \emph{\bibinfo{title}{Geometric Algebra for Physicists}}
  (\bibinfo{publisher}{Cambridge University Press}, \bibinfo{year}{2003}).

\end{thebibliography}

\end{document}